\documentclass[]{aa}
\usepackage{times}
\usepackage{graphics}

\begin{document}

   \thesaurus{07          
              (02.03.3;   
               02.08.1;   
               02.12.1;   
               06.01.1;   
               06.07.2;   
               06.16.2)}   

   \title{The effects of numerical resolution on hydrodynamical 
surface convection simulations and spectral line formation}

   \author{M. Asplund\inst{1,2}, H.-G. Ludwig\inst{3,4}, 
          \AA. Nordlund\inst{3} and R.F. Stein\inst{5}
          }

   \offprints{Martin Asplund (martin@astro.uu.se)}

   \institute{
              NORDITA, 
              Blegdamsvej 17, 
              DK-2100 ~Copenhagen {\O}, 
              Denmark         
              \and
              present address: Uppsala Astronomical Observatory,
              Box 515,
              SE-751 20 ~Uppsala,
              Sweden
              \and
              Astronomical Observatory, NBIfAFG, 
              Juliane Maries Vej 30,
              DK-2100 ~Copenhagen \O, 
              Denmark
              \and
              present address: ENS-CRAL,
              46 all\'ee d'Italie,
              F-69364 Lyon Cedex 07,
              France
              \and
              Dept. of Physics and Astronomy, 
              Michigan State University, 
              East Lansing, MI 48823, USA
              }

   \date{Received: March 3, 2000; accepted: May 4, 2000 }

\authorrunning{Asplund et al.}
\titlerunning{Resolution effects in convection simulations
and line transfer}

\maketitle

   \begin{abstract}

The computationally demanding nature of radiative-hydrodynamical 
simulations of stellar surface convection warrants
an investigation of the sensitivity of the convective
structure and spectral synthesis to the numerical resolution
and dimension of the simulations, which is presented here.

With too coarse a resolution the predicted spectral lines tend
to be too narrow, reflecting insufficient Doppler 
broadening from the convective motions, while at the currently
highest affordable resolution 
the line shapes have converged essentially perfectly to the
observed profiles. Similar conclusions are drawn from the
line asymmetries and shifts. 
Due to the robustness of the pressure and temperature structures
with respect to the numerical resolution, strong Fe lines 
with pronounced damping wings and H lines are essentially immune
to resolution effects, and can therefore be used for improved 
$T_{\rm eff}$ and log\,$g$ determinations even at very modest 
resolutions.
In terms of abundances, weak Fe\,{\sc i} and Fe\,{\sc ii} lines
show a very small dependence ($\simeq 0.02$\,dex) while for intermediate
strong lines with significant non-thermal broadening the sensitivity
increases ($\la 0.10$\,dex). 

Problems arise when using 2D convection simulations to describe an
inherent 3D phenomenon, which translates to inaccurate atmospheric 
velocity fields and temperature and pressure structures.
In 2D the theoretical line profiles tend to be too shallow and broad
compared with the 3D calculations and observations, in particular for
intermediate strong lines. In terms of abundances, the 2D results are
systematically about 0.1\,dex lower than for the 3D case for Fe\,{\sc i} lines. 
Furthermore, the predicted line asymmetries
and shifts are much inferior in 2D with discrepancies amounting to
$\sim 200$\,m\,s$^{-1}$. 
Given these shortcomings and computing time 
considerations it is better to use 3D simulations of even modest resolution
than high-resolution 2D simulations.

      \keywords{Convection -- Hydrodynamics --  
                Line: formation -- Sun: abundances --
                Sun: granulation -- Sun: photosphere 
               }
   \end{abstract}

\section{Introduction}

Recent progress in hydrodynamical simulations of solar and stellar
granulation and spectral line transfer 
(e.g. Nordlund \& Dravins 1990; Stein \& Nordlund 1989, 1998;
Asplund et al. 1999, 2000a,b, hereafter Paper I and II)
have opened the door for more secure analyses of observed stellar
spectra. Due to the parameter-free nature of such convection
simulations it is now possible to compute self-consistent 3-dimensional
(3D) model atmospheres and line profiles without relying on 
various ad-hoc parameters, such as mixing length
parameters, micro- and macroturbulence, which are necessary in 1D analyses.
Furthermore, the simulations successfully reproduce a variety
of observational diagnostics like the solar granulation geometry,
flow and brightness properties,  helioseismological constraints, and
spectral line shapes, asymmetries and shifts
(e.g. Stein \& Nordlund 1998; Rosenthal et al. 1999; 
Georgobiani et al. 2000; Paper I and II). 

The advantages of using 3D convection simulations are therefore clear
and the results in terms of e.g. derived stellar elemental abundances
should be more reliable, besides of course the fact that the simulations 
will themselves shed further light on the nature of stellar convection. 
The major drawback with such a procedure is similarly obvious.
The required computing time to perform even a single solar surface
convection simulation sequence of one solar hour ($\approx 10^4$ Courant time steps)
at the current highest numerical resolution (200\,x\,200\,x\,82) is today
typically two CPU-weeks on a supercomputer such as the Fujitsu VX-1
(peak speed: 2.2\,Gflops/CPU).
Additionally, the memory requirement is about 100\,b/gridpoint  (i.e. 320\,Mb
for the same mesh) and the output consists of about 4\,Gb of data.
Furthermore, the 3D spectral line transfer is
also relatively time-consuming even with the much simplifying
assumption of LTE; typically the equivalent of $> 10^5$
1D spectral synthesis calculations are necessary to obtain 
statistically significant spatially and temporally averaged line
profiles (e.g. Paper I and II). 

Any approach that can significantly reduce the demand on
computing power is therefore worthwhile to pursue. An obvious solution 
to the problem is to settle for poorer numerical resolution
in the 3D simulations or even restrict the computations to 2D,
since the CPU-time scales roughly with the number of
grid points. Depending on ones ultimate goal such a procedure may
or may not be acceptable. To exemplify, a more demanding resolution
is likely necessary if one is interested in studying detailed
line asymmetries with a 10\,m\,s$^{-1}$ accuracy or
small-scale dynamic phenomena in the photosphere than if one is
interested in the required entropy jump between the surface
and the interior adiabatic structure in order to calibrate 1D stellar evolution
models (Ludwig et al. 1999). 

Here we present an investigation of the influence of the resolution
and dimension of the convection simulations on stellar spectroscopy.
In particular we will focus on the resulting spatially and 
temporally averaged line profiles, shifts and asymmetries, although 
the differences in atmospheric temperature and velocity structures
will also be discussed as they contain the keys how to interpret
the spectral variations.
For the purpose we have concentrated on Fe and H lines due to
their wide applicability and diagnostic abilities for 
stellar astrophysics. The comparisons between 3D simulations of
different resolutions and between 2D and 3D simulations have
been performed strictly differentially in the sense that the 
differences are restricted only to the resolution and dimension
of the simulations with all other numerical details being identical. 
The basic question we attempt to address is therefore: What is the
required minimum resolution in order to still obtain realistic
results?

\section{Hydrodynamical surface convection simulations
\label{s:simulations}}

The 3D and 2D model atmospheres of the solar granulation which form the
basis of the present investigation have been obtained with a time-dependent,
compressible, radiative-hydrodynamics code developed to study solar
and stellar granulation (e.g. Nordlund \& Stein 1990; 
Stein \& Nordlund 1989, 1998; Asplund et al. 1999; Ludwig et al. 1999; Paper I and II). 
The code solves the hydrodynamical
conservation equations of mass, momentum and energy together with the
3D or 2D equation of radiative transfer under the assumption of LTE
and the opacity binning technique (Nordlund 1982). For further details on
the numerical algorithms the reader is referred to Stein \& Nordlund (1998).

To study the effects of numerical resolution in 
3D solar convection simulations and spectral synthesis, non-staggered
Eulerian meshes with 200\,x\,200\,x\,82, 100\,x\,100\,x\,82, 50\,x\,50\,x\,82
and 50\,x\,50\,x\,63 gridpoints have been used. In all other respects the
simulations are identical. The depth scale ranges from 1.0\,Mm above 
to 3.0\,Mm below the visual surface and has been optimized to provide
the best resolution in the layers with the strongest variations of
d$T$/dz and d$^2T$/dz$^2$, which for the Sun
occurs around $z = 0$\,Mm. All three simulations with 82 depth points have
furthermore identical vertical depth scales. 
In all cases the total horizontal dimension measures
6.0\,x\,6.0\,Mm, which is sufficiently large to include $\ga 10$ granules
at any time. The equation-of-state was provided by Mihalas et al. (1988)
with a standard solar chemical composition (Grevesse \& Sauval 1998).
The radiative transfer during the convection simulations included
up-to-date continuous (Gustafsson et al. 1975 with subsequent updates)
and line opacities (Kurucz 1993) and was solved for eight inclined rays 
(2 $\mu$-angles and 4 $\varphi$-angles).
The simulations cover the same 50\,min time-sequence of the solar granulation with
the initial snapshot for the 100\,x\,100\,x\,82, 50\,x\,50\,x\,82 and 
50\,x\,50\,x\,63 cases interpolated from the 200\,x\,200\,x\,82 simulation. 
The resulting effective temperatures $T_{\rm eff}$
\footnote{Since the entropy of the inflowing gas at the lower boundary is used
as a boundary condition rather than the emergent luminosity at the surface
as commonly done in classical 1D model atmospheres, the resulting 
$T_{\rm eff}$ varies slightly with time throughout the simulation.}
are therefore very similar in all cases: 
$T_{\rm eff} = 5767\pm21$, $5768\pm21$, $5768\pm18$\,K and $5774\pm14$\,K,
in order of decreasing resolution.
The larger time variation
in $T_{\rm eff}$ with increasing resolution appears to
be significant. Provisionally we attribute it to the better
ability to describe small scale events such as edge brightening of
granules (Stein \& Nordlund 1998) with an improved resolution. 

To investigate the effects of dimension in the convection simulations
and spectral synthesis, we have additionally performed a similar 2D solar
convection simulation with a numerical grid of 100\,x\,82. 
The horizontal and vertical extensions are identical to those of the
3D simulations, as are the input physics in terms of equation-of-state,
opacities and chemical composition. 
In order to achieve the correct $T_{\rm eff}$ the entropy of the inflowing
gas at the lower boundary had to be adjusted compared with the 3D simulations.
The difference in inflowing entropy amounts to $4.2 \cdot 10^7$\,erg/g/K,
which is consistent with the findings of Ludwig et al. (in 
preparation) who used similar convection simulations.
The 2D simulation sequence used for the subsequent spectral synthesis
covers in total 16.5\,hr solar time, although the
full simulation is significantly longer (23\,hr); the first part of the 
simulation has not been used in order to allow the convective structure to
achieve thermal relaxation after modifying the entropy structure of the initial
vertical slice which was taken from a 3D snapshot.
The resulting $T_{\rm eff}$ is similar for the two corresponding 
simulation sequences used
for the line formation calculations:
$T_{\rm eff} = 5732 \pm 87$\,K (2D) and 
$T_{\rm eff} = 5768 \pm 21$\,K (3D: 100\,x\,100\,x\,82); 
the difference of 36\,K has a minor
influence on the resulting line profiles. 
The larger time variability of the emergent luminosity in 2D
is a natural consequence of the smaller area coverage, which makes the influence
from individual granules more pronounced.

\section{Spectral line calculations and observational data}

Prior to the 3D and 2D spectral syntheses the convection simulations
were interpolated to a finer depth scale to improve the accuracy 
of the radiative transfer. The new vertical scale only extended 
down to 0.7\,Mm instead of 3.0\,Mm for the original solar simulations.
Additionally the horizontal resolutions
of the original simulation sequences were in the case of the 100\,x\,100\,x\,82
and 200\,x\,200\,x\,82 simulations decreased to 50\,x\,50\,x\,82 
while retaining the physical horizontal dimensions of the numerical box
to ease the computational burden. Various test calculations assured 
that the procedure did not introduce any differences in the
resulting line profiles or bisectors. 
It is important to emphasize that the effects of the numerical
resolution in the convection simulations on e.g. the atmospheric temperature
and velocity structures will be fully contained in the interpolated
snapshots used for the line transfer calculations; in the end the
spatially and temporally averaged profiles are averaged over 
sufficiently many columns (about 250\,000 comparative 1D model atmospheres) 
that 4 or 16 times more columns will not make any noticeable difference.
The theoretical line formation was performed for each simulation snapshot 
(30\,s intervals) of the full convection sequences 
described in Sect. \ref{s:simulations}
covering in total 50\,min (3D) and 16.5\,hrs (2D) solar time, i.e.
100 and 1980, respectively, different snapshots.

For the spectral syntheses the assumption of LTE ($S_\nu = B_\nu$)
has been made throughout. The background opacities were computed using the
Uppsala opacity package (Gustafsson et al. 1975 with subsequent 
updates) while the equation-of-state was supplied by 
Mihalas et al. (1988). Here only intensity
profiles at disk-center have been calculated. 

\begin{table}[t!]
\caption{The continuum intensity contrast for simulations of
different resolution and dimension
\label{t:irms}
}
\begin{tabular}{lccc} 
 \hline
Simulation &  \multicolumn{3}{c}{$I_{\rm rms}/I$ @ 620\,nm} \\
\cline{2-4} 
           & No smearing & Telescope$^{\rm a}$ & Telescope+seeing$^{\rm b}$ \\
\hline 

$200^2{\rm x}82$ & $0.168\pm0.004$ & $0.142\pm0.004$ & $0.086\pm0.003$ \\  
$100^2{\rm x}82$ & $0.166\pm0.005$ & $0.144\pm0.005$ & $0.089\pm0.004$ \\ 
$50^2{\rm x}82$  & $0.164\pm0.007$ & $0.146\pm0.007$ & $0.096\pm0.006$ \\ 
$50^2{\rm x}63$  & $0.157\pm0.007$ & $0.140\pm0.006$ & $0.092\pm0.005$ \\ 
$100{\rm x}82$   & $0.175\pm0.031$ & $0.162\pm0.030$ & $0.119\pm0.028$ \\

\hline 
\end{tabular}
\begin{list}{}{}
\item[$^{\rm a}$] One Lorentzian of width a=0.15\,Mm (Nordlund 1984)
\item[$^{\rm b}$] Two Lorentzians of widths a=0.15\,Mm and b=1.5\,Mm with 
weighting p=0.4 (Nordlund 1984) 
\end{list}

\end{table}

The sample of lines used in the present project consisted of
the weak and intermediate strong lines ($W_\lambda \le 10$\,pm)
Fe\,{\sc i} and Fe\,{\sc ii} lines of Blackwell et al. (1995)
and Hannaford et al. (1992), respectively. Additionally five
strong Fe\,{\sc i} lines (407.2, 491.9, 523.3, 526.9 and 544.7\,nm)  
with pronounced damping wings were included with parameters identical
to those adopted in Paper II. For the Fe\,{\sc i} lines the collisional broadening 
was included following the recipes of Anstee \& O'Mara (1991, 1995),
Barklem \& O'Mara (1997), Barklem et al. (1998, 2000b), 
while for the Fe\,{\sc ii} lines the classical damping result was adopted but
enhanced with a factor $E=2$. The lines and their atomic
data are listed in Table \ref{t:lines}. In particular, the
accurate laboratory wavelengths of Nave et al. (1994) and 
S. Johansson (Lund, 1998, private communication) for 
Fe\,{\sc i} and Fe\,{\sc ii} have been adopted. 
Finally, H$\alpha$ and H$\beta$ profiles have been calculated using
the Vidal et al. (1973) broadening theory. 

Since the resulting line profiles turned out to depend on 
the numerical resolution (Sect. \ref{s:shape3d} and \ref{s:shape2d})
we had to resort to the use of equivalent widths when comparing the derived 
Fe abundances for the weak and intermediate strong lines, although
this is not necessary with the highest resolution (Paper I and II).
For the purpose we adopted the values used by Blackwell et al. (1995)
and Hannaford et al. (1992) for the Fe\,{\sc i} and Fe\,{\sc ii} lines,
respectively. As in Paper I and II, 
each theoretical line profile was computed with three different
elemental abundances ($\Delta {\rm log} \epsilon_{\rm Fe} = 0.2$\,dex)
from which the abundance returning the observed equivalent width was 
interpolated. When comparing the synthesized profiles and bisectors
with observations the solar intensity atlas of 
Brault \& Neckel (1987) and Neckel (1999), which
is based on an accurate wavelength calibration
(Allende Prieto \&  Garc\'{\i}a L{\'o}pez 1998), has been utilized. 
The same measured line bisectors as presented in Paper I
have been adopted here. 

\section{Effects of resolution in 3D}

\begin{figure}[t]
\resizebox{\hsize}{!}{\includegraphics{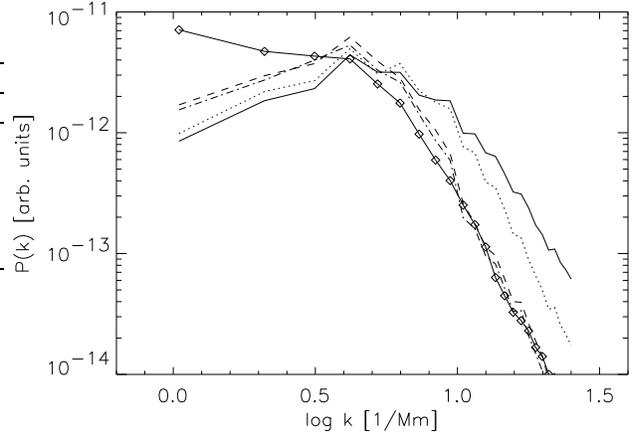}}
\caption{The average power spectrum of the continuum intensity at
620\,nm for different resolutions of the solar convection simulation:
200\,x\,200\,x\,82 (solid line), 100\,x\,100\,x\,82 (dotted line),
50\,x\,50\,x\,82 (dashed line), 50\,x\,50\,x\,63 (dot-dashed line) and
100\,x\,82 (solid line with diamonds). No smearing representing
the finite telescope and atmospheric seeing has been applied to the
intensity images from the granulation simulations. Note that 
the higher resolution simulations resolve smaller scales than shown in
the figure but those have been omitted here to correspond directly with the
lowest horizontal resolution studied (50\,x\,50)
}
         \label{f:intpower}
\end{figure}

\begin{figure}[t]
\resizebox{\hsize}{!}{\includegraphics{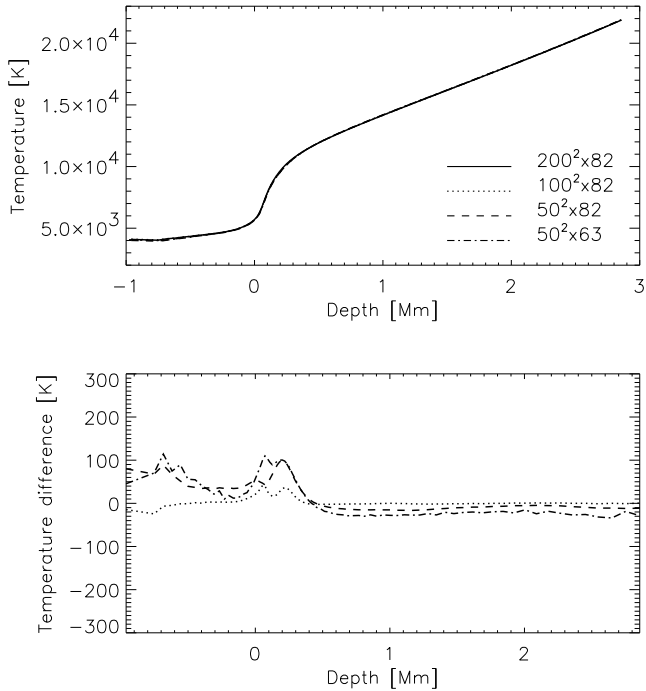}}
\caption{{\it Upper panel:}
The mean temperature structure in solar convection simulations of 
different resolution:  200\,x\,200\,x\,82 (solid), 100\,x\,100\,x\,82 (dotted),
50\,x\,50\,x\,82 (dashed) and 50\,x\,50\,x\,63 (dot-dashed).
{\it Lower panel:} The temperature differences relative to the 
200\,x\,200\,x\,82 simulation (positive values correspond to lower
temperatures than for 200\,x\,200\,x\,82).
The mean temperature is a very robust property with
no significant difference between the simulations
}
         \label{f:temp3d}
\end{figure}

\subsection{Effects on continuum intensity contrast \label{s:irms3d}}

The horizontal temperature variations in the photosphere
between granules and intergranular lanes produce a pronounced 
continuum intensity contrast, which together with the velocity field
modify the resulting spectral line shapes and asymmetries.
Provided the image degradation introduced by the finite resolution
of the telescope, instrumentation and telluric atmosphere can be
modelled, the observed
intensity contrast of the solar granulation can thus function as an 
important additional test to confront with predictions from 
surface convection simulations. 
Table \ref{t:irms} summarizes the
theoretical results calculated at $\lambda = 620$\,nm 
from a temporal average of simulations of different
numerical resolution, both from the raw unsmeared images and when
accounting for the seeing. In 3D, $I_{\rm rms}$ varies only slightly
with resolution, being 16.8\% at the highest resolution without any
smoothing. The slight increase with improved resolution follows from the lower
effective numerical viscosity in the simulations, which allows more
power on smaller spatial scales, as evident from Fig. \ref{f:intpower}.
It should be noted that the intensity contrast is much smaller than the
temperature contrast in the photosphere, since the surface with 
continuum optical depth unity is corrugated and thus the 
high temperature gas is partly hidden from sight (Stein \& Nordlund 1998). 

The effects of seeing is clearly visible in Table \ref{t:irms}, 
bringing $I_{\rm rms}$ down to about 8-9\%. Although poorly known 
the atmospheric and telescope point spread functions (PSF) have 
here been provisionally accounted for by convolving the original 
intensity images with two Lorentz profiles with widths $a=0.15$\,Mm 
and $b=1.5$\,Mm and relative weighting of $p=0.4$, following the 
discussion in Nordlund (1984). Physically, the narrow component 
corresponds to the broadening by the telescope and the extended 
part scattering by the atmospheric seeing. The latter profile is
the most crucial factor when comparing with observations but also 
the more uncertain. Quantitatively different contrasts are obtained 
when adopting different broadening parameters or using Gaussian 
PSFs rather than Lorentz profiles. The reversal of the trend with 
numerical resolution is due to the increased power at smaller 
spatial scales with improved resolution (Fig. \ref{f:intpower}) 
which makes the contrast slightly smaller at the relevant larger 
scales after smearing.  A detailed comparison with observations 
would require an accurate modelling of the convolution functions,
which is beyond the scope of the present investigation. We happily 
note though that the values given in Table \ref{t:irms} are in 
close agreement with the best solar images (e.g. Lites et al. 
1991), which suggest $I_{\rm rms} = 8-9$\% at 620\,nm.

It should be noted, however, that due to the obscuration from the poorly
understood seeing effects, probably the best test of the intensity 
contrast comes from a comparison of spectral line asymmetries rather than
images. Spatially averaged bisectors 
are seeing-independent measures of the product of
$I_{\rm rms}$ and $v_{\rm z,rms}$ but the latter is fixed by the
width of the spectral lines. The excellent agreement between predicted
and observed line shapes and asymmetries indirectly also show that the
theoretical $I_{\rm rms}$ must be very close to the real solar value.

\begin{figure}[t]
\resizebox{\hsize}{!}{\includegraphics{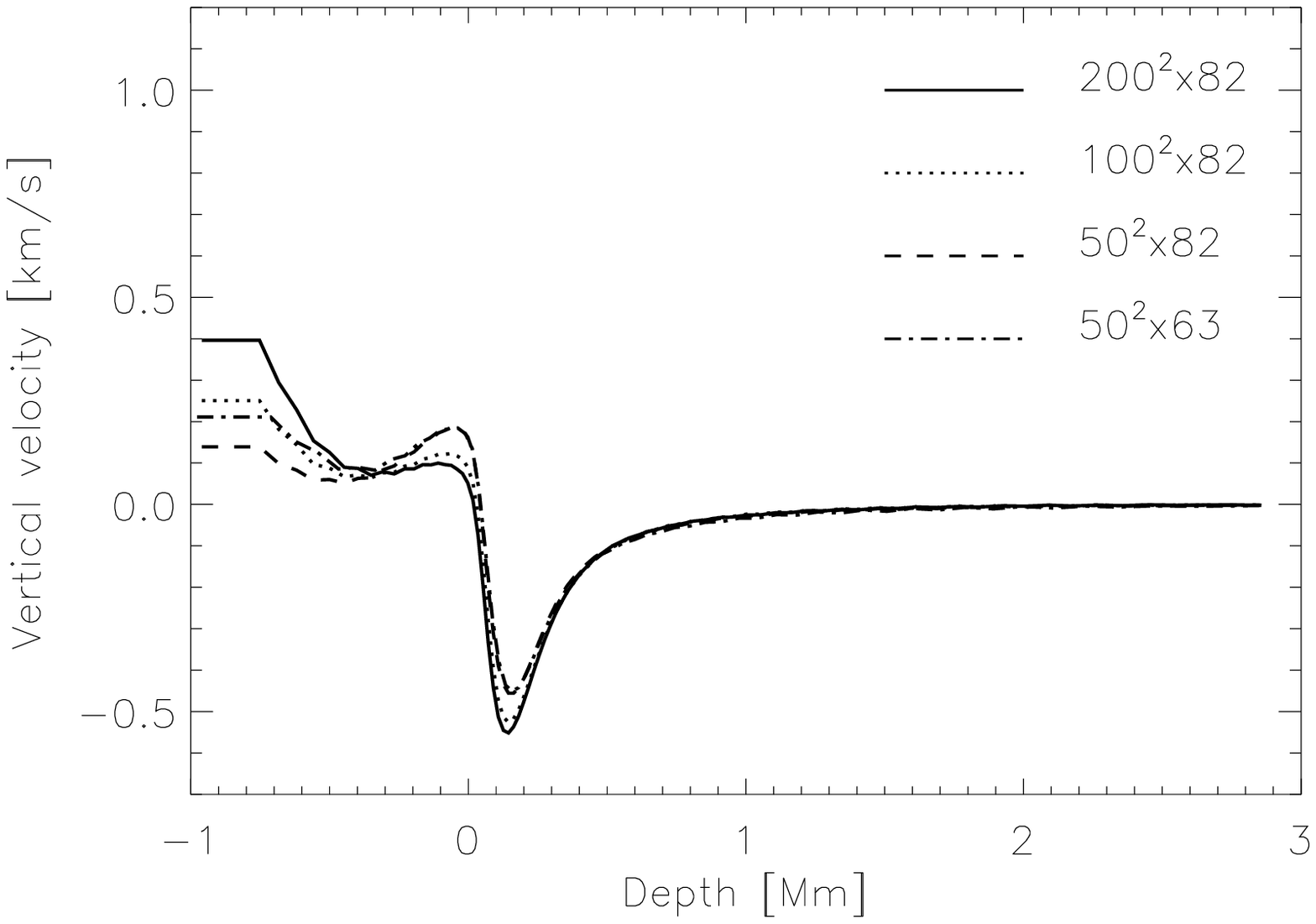}}
\resizebox{\hsize}{!}{\includegraphics{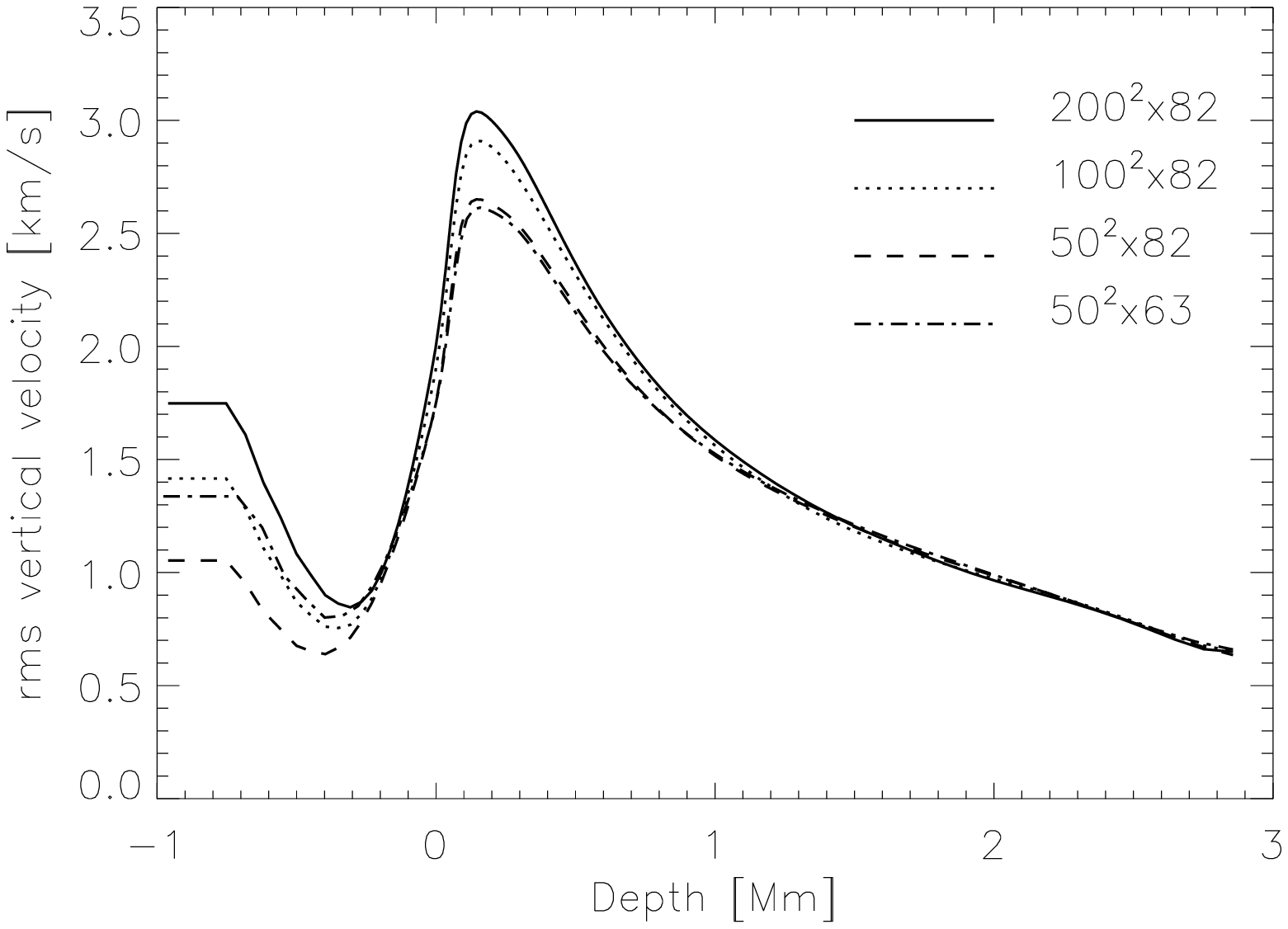}}
\caption{The temporal mean vertical velocity ({\it upper panel}) and 
horizontal rms vertical velocity ({\it lower panel}) 
in solar convection simulations of 
different resolution:  
200\,x\,200\,x\,82 (solid), 100\,x\,100\,x\,82 (dotted), 
50\,x\,50\,x\,82 (dashed) and 50\,x\,50\,x\,63 (dot-dashed). 
The upper boundary condition, which stipulates equal vertical
velocities in the two uppermost layers, causes the horizontal 
part of the curves at the top. 
Positive vertical velocities correspond to downflows
}
         \label{f:vz3d}
\end{figure}

\begin{figure}[t]
\resizebox{\hsize}{!}{\includegraphics{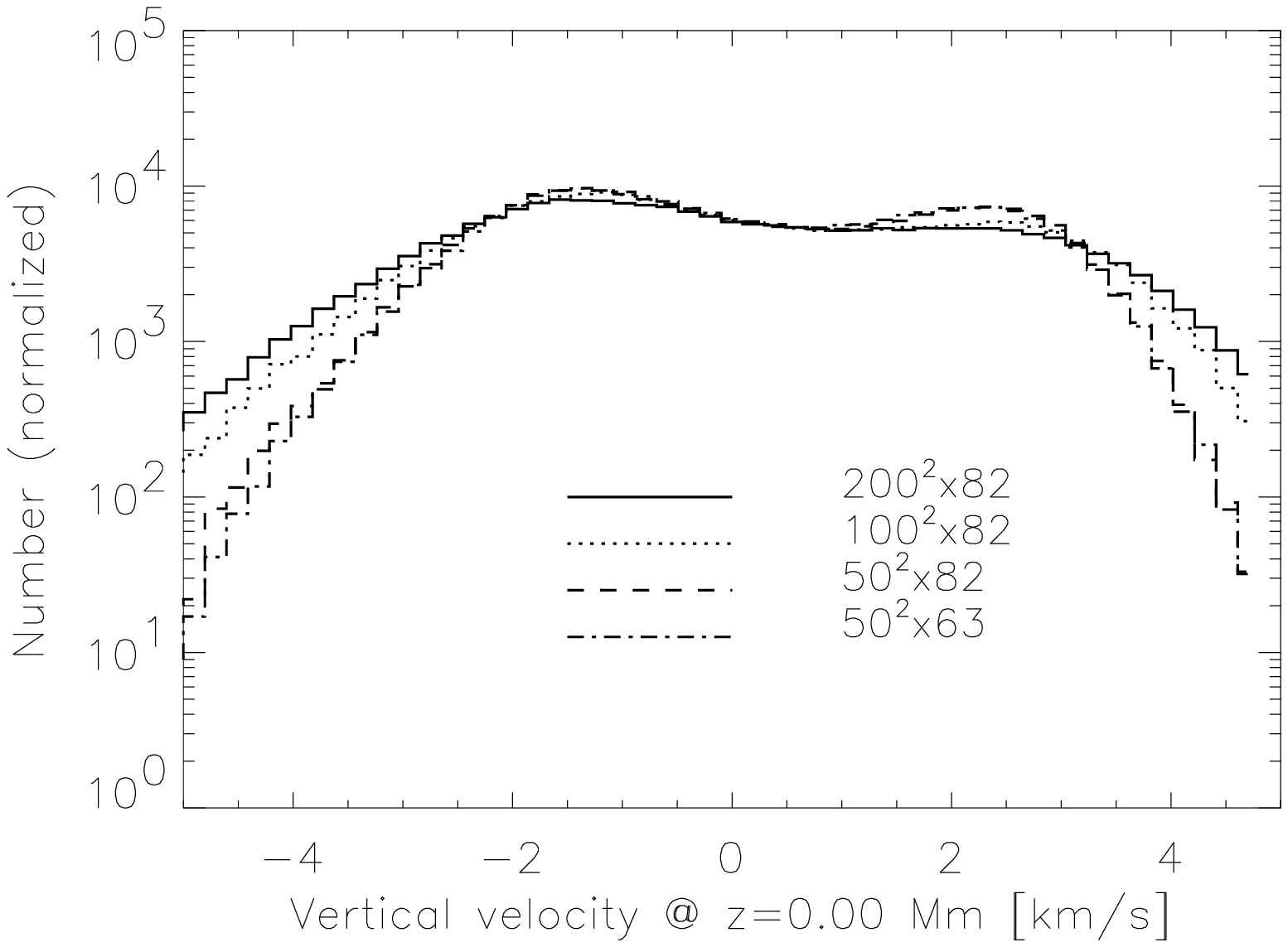}}
\resizebox{\hsize}{!}{\includegraphics{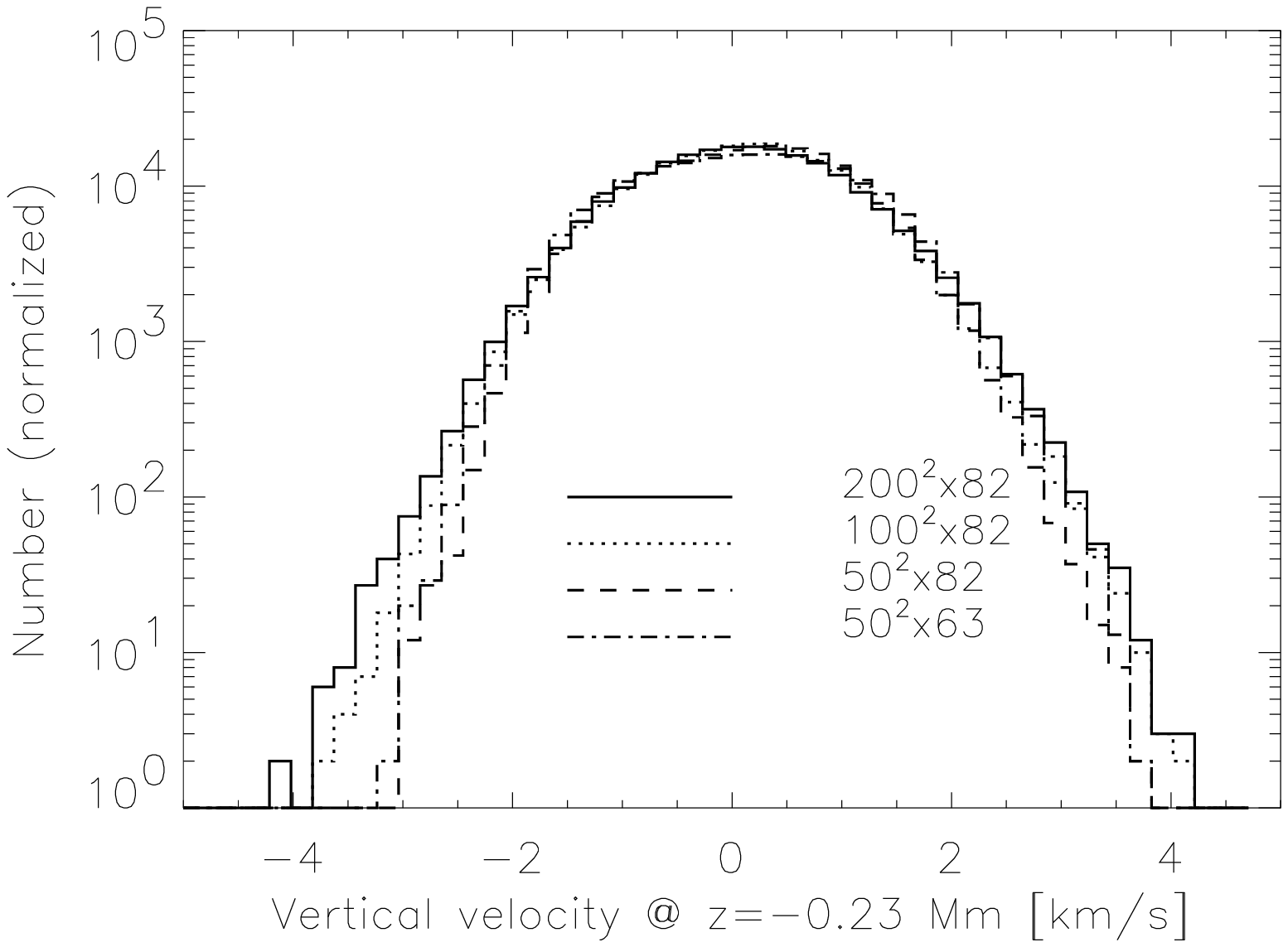}}
\caption{Histogram of the vertical velocity distribution at the
visible surface ($z=0.0\,$Mm, {\it upper panel}) and  
in higher line-forming layers ($z=-0.23\,$Mm, {\it lower panel})
in the solar convection simulations at different numerical resolution. 
Positive vertical velocities correspond to downflows. 
With a higher resolution the tails of the velocity distributions are better
sampled
}
         \label{f:vzdist3d}
\end{figure}

\subsection{Effects on line shapes and abundances \label{s:shape3d}}
 
\begin{figure}[t]
\resizebox{\hsize}{!}{\includegraphics{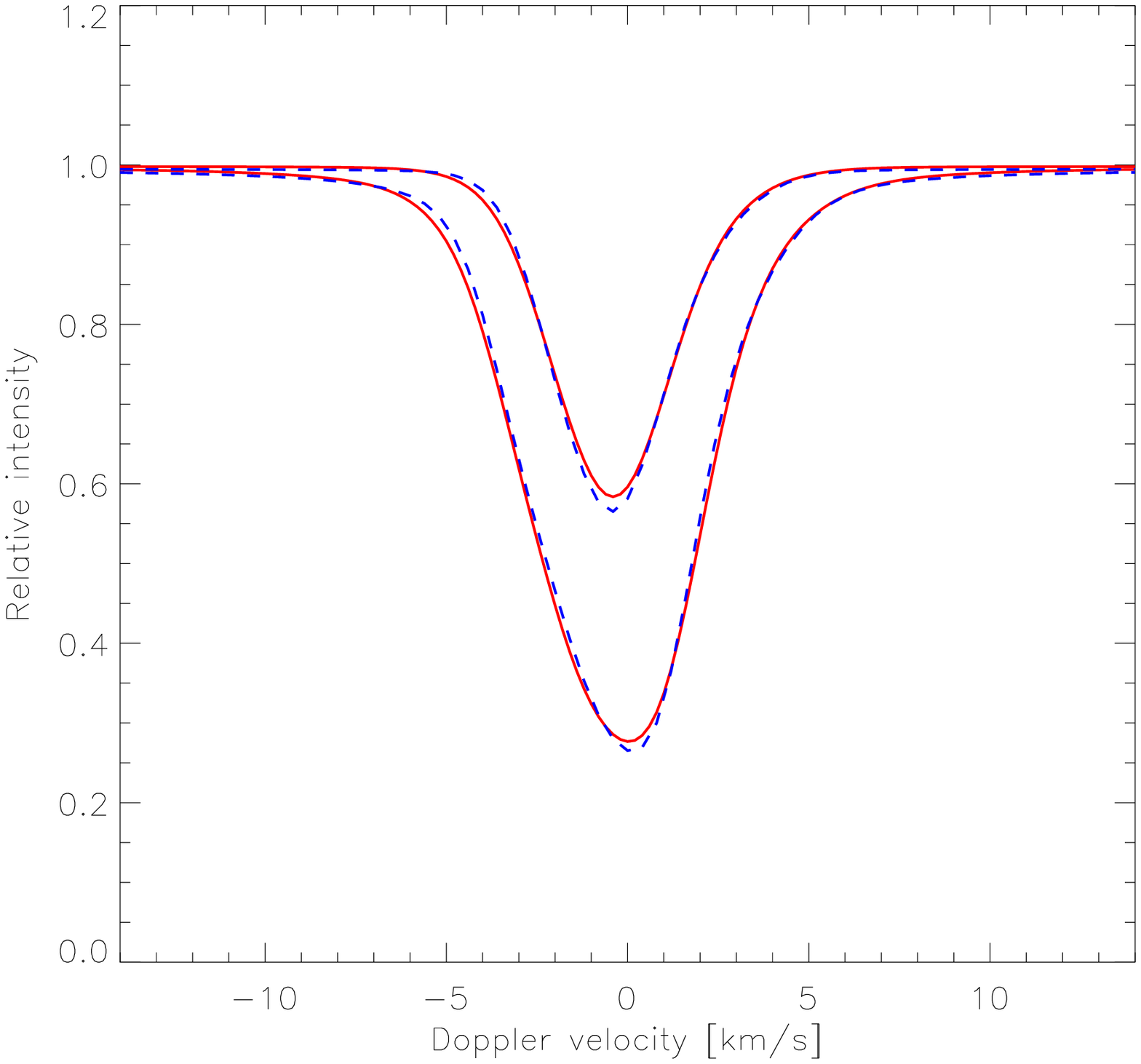}}
\caption{The predicted Fe\,{\sc i} 608.2 (weaker) and 621.9\,nm (stronger) 
lines at two different resolutions of the solar convection simulation:
200\,x\,200\,x\,82 (solid) and 50\,x\,50\,x\,82 (dashed). 
All profiles have been computed with 
log$\,\epsilon_{\rm Fe}=7.50$}
         \label{f:prof3d}
\end{figure}

The non-thermal broadening of spectral lines stems predominantly from
the Doppler shifts due to the convective flows inherent in
the granulation (e.g. Paper I).  
A substantial part of the total line strength of solar Fe lines 
is therefore contributed by the convective broadening, 
in particular for stronger, partly saturated, lines. 
The mean temperature structures in the simulations are essentially 
determined by mass conservation and the amount of radiative cooling 
at the visible surface, which should be
basically independent of the numerical resolution (Stein \& Nordlund 1998).
As seen in Fig. \ref{f:temp3d} the temperature structures in the various
3D simulations are indeed very similar, which is also true for the
average entropy structures. From this follows that the {\em average} vertical
velocity is similar in the convection zone (Fig. \ref{f:vz3d})
but it is not true that also the velocity
distributions will be insensitive to the resolution (Stein \& Nordlund 1998).
\footnote{In the higher atmospheric layers, the mean vertical velocities reflect
more propagating waves than convective motions, which are 
excited by small-scale acoustic events and therefore show a
sensitivity to the resolution.}
In particular this is not correct in the photospheric layers which are important
for the line broadening. Although the bulk of the velocity distributions
only show minor differences, the high-velocity tails of the distributions
are indeed quite sensitive to the resolution (Fig. \ref{f:vzdist3d}). 
With a better resolution
the more extreme velocities, which can contribute significant line
broadening, are more likely due to the smaller effective viscosity. 
On the basis of such considerations, 
one would therefore expect the widths and strengths of weak lines 
to be less sensitive to the resolution while stronger, partly
saturated, lines should show a higher degree of dependence. 
It can be noted that the vertical velocity distributions are quite symmetric
in terms of up- and downflows at the visible surface and line-forming regions,
while it is markedly skewed in the deeper layers, with much larger descending
velocities (Stein \& Nordlund 1998). In warmer stars such as F stars 
with more naked granulation (Nordlund \& Dravins 1990; Asplund et al., in 
preparation) the asymmetric velocity distributions also in the photospheric
layers will be manifested in the line formation. 

\begin{figure*}[t]
\resizebox{\hsize}{!}{\includegraphics{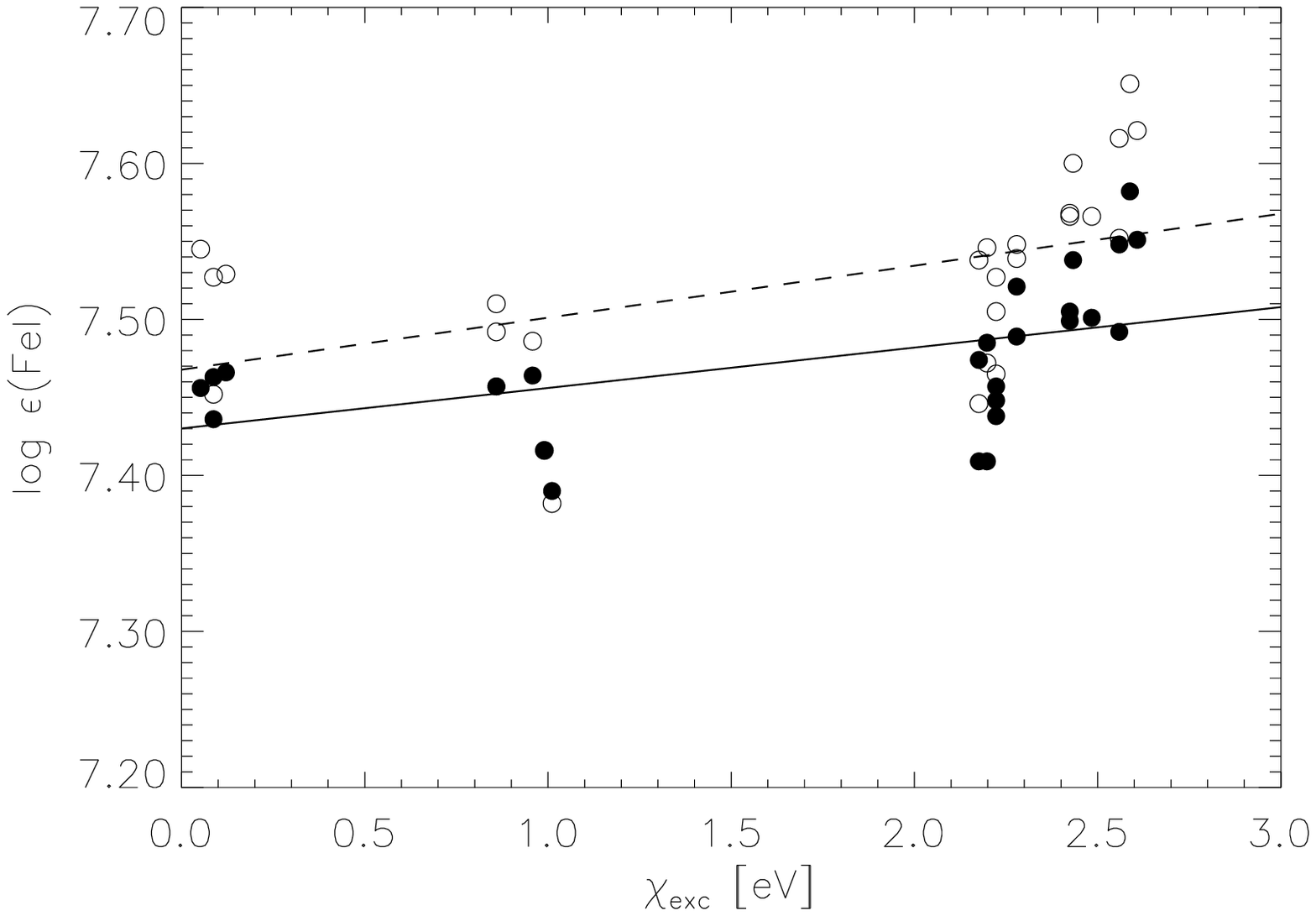}
                      \includegraphics{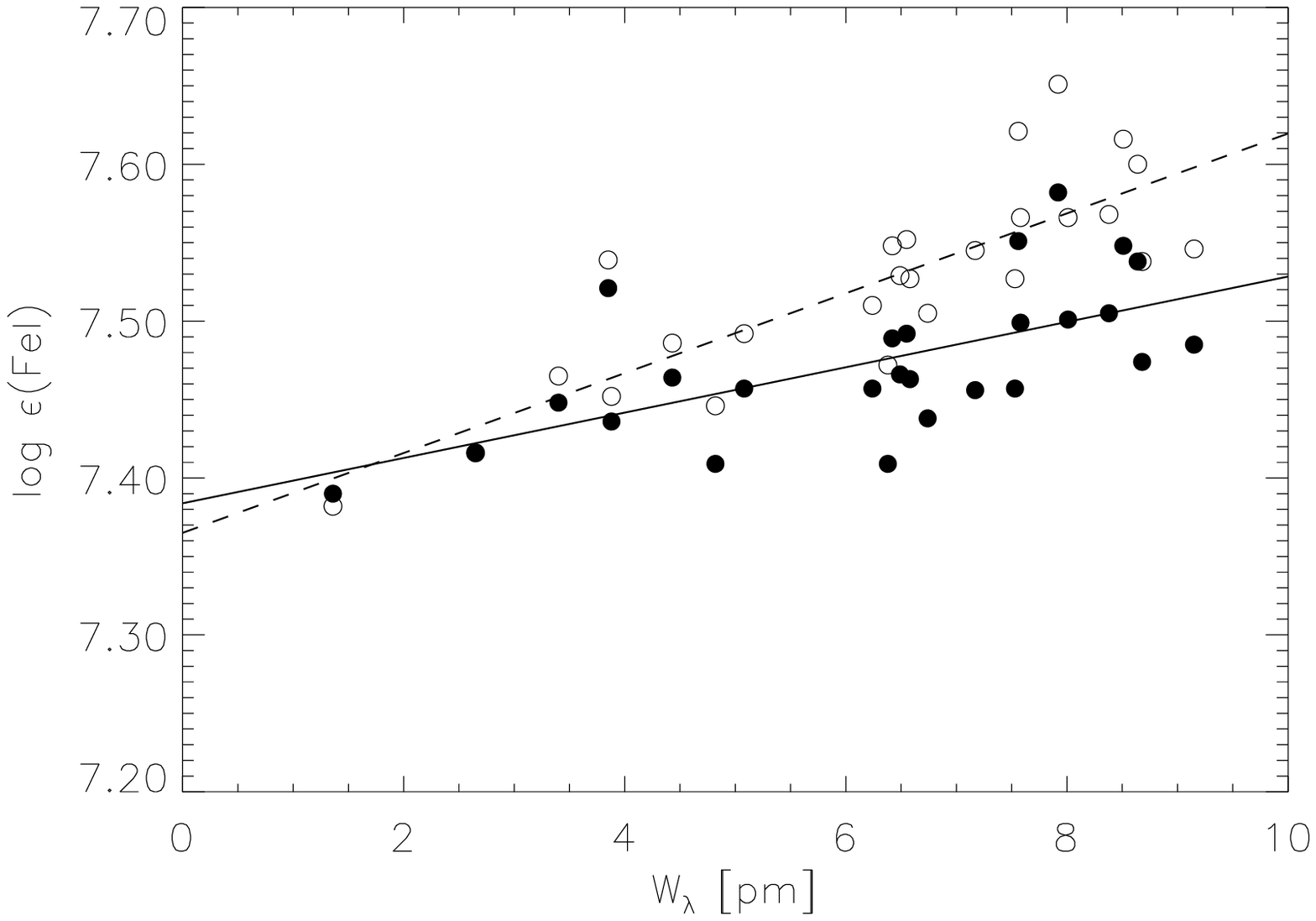}}
\resizebox{\hsize}{!}{\includegraphics{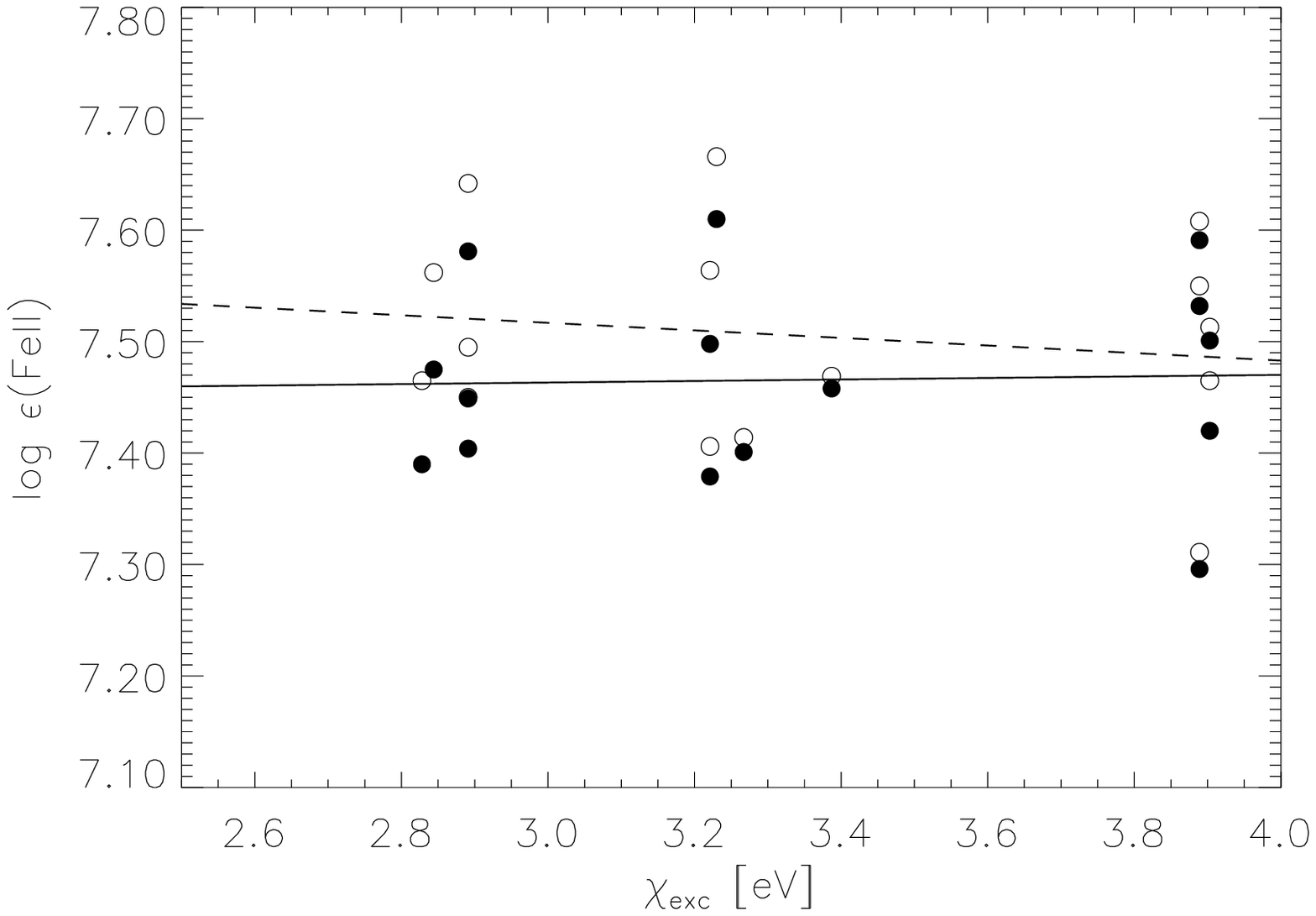}
                      \includegraphics{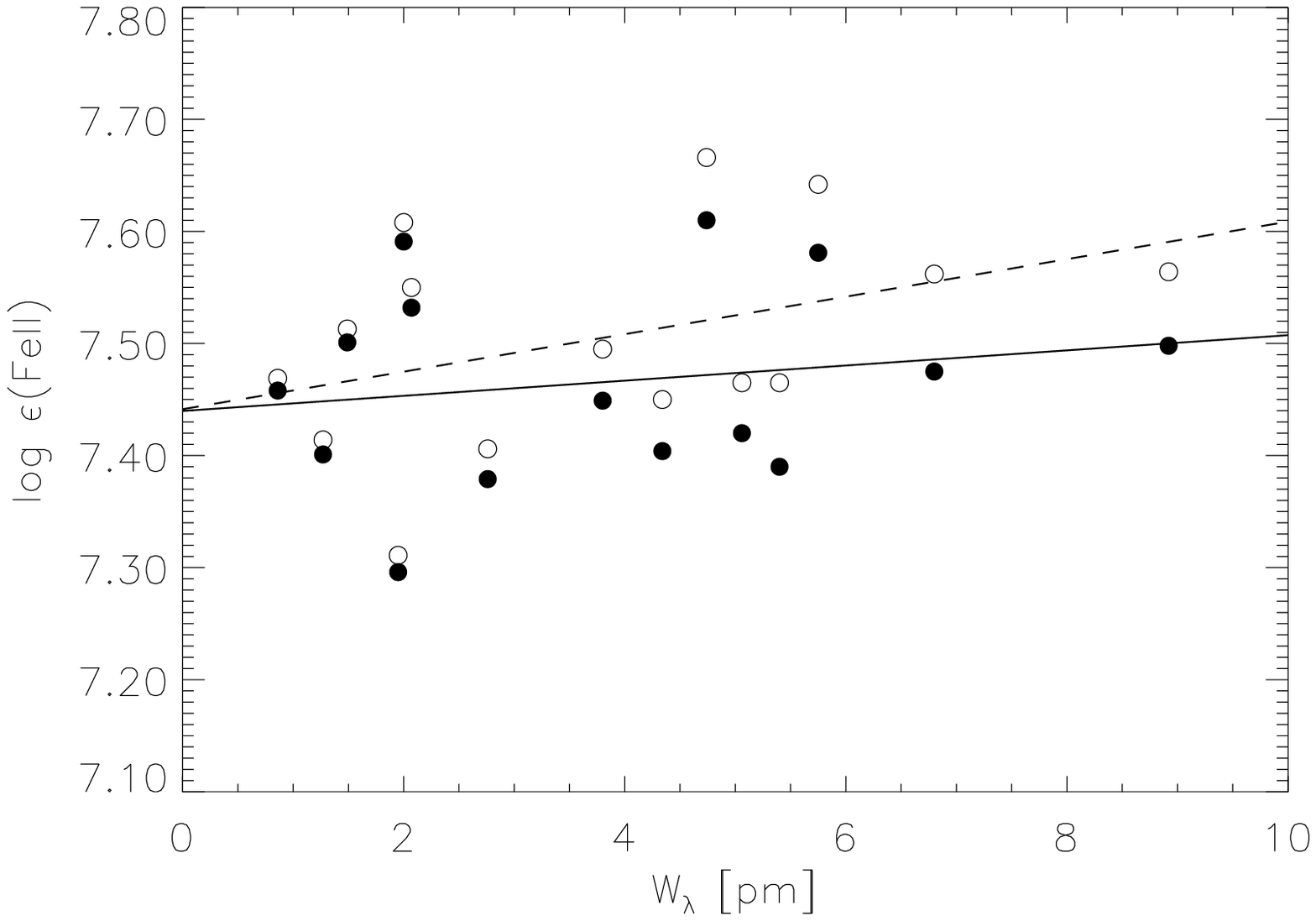}}
\caption{The abundances derived from Fe\,{\sc i} ({\it upper panel}) and 
Fe\,{\sc ii} ({\it lower panel}) lines as a function of the 
excitation potential ({\it left panel}) and equivalent widths 
({\it right panel}) for two different resolutions:
200\,x\,200\,x\,82 (filled circles) and
50\,x\,50\,x\,63 (open circles). 
The solid and dashed lines are least-square fits
to the two sets of abundances. The abundances for the intermediate resolutions
are not shown for clarity but generally fall in between these two cases. Note
that the trends are more pronounced here than in Paper II due to the
use of equivalent widths instead of profile fitting
}
         \label{f:abu3d}
\end{figure*}

As suspected the line widths for a given line strength increase
with increasing numerical resolution, which is illustrated in Fig. \ref{f:prof3d}.
Since at the highest resolution the line profiles match almost perfectly
the observed profiles (Paper I and II; Asplund 2000), with a lower resolution
the computed lines tend to be slightly too deep and narrow for a given
equivalent width. 
Somewhat surprisingly it was found that the additional broadening
preferentially was contributed to the blue wing of the profiles. 
It appears that the increased number of regions with downflow 
velocities of $\approx 2.5$\,km\,s$^{-1}$ with poorer resolutions
largely compensate the lack of even larger redshifted velocities. 

Although the lower resolution simulations do not provide all the necessary
line broadening, one may hope to be able to derive elemental abundances
from equivalent widths even with a limited resolution. 
Table \ref{t:lines} lists the individual
abundances thus derived for the different simulations. 
The mean Fe abundances from the weak and intermediate strong 
Fe\,{\sc i} lines are $7.48\pm0.05, 7.51\pm0.05$, $7.52\pm0.06$
and $7.53\pm0.06$, in order of decreasing
numerical resolution, while for the Fe\,{\sc ii} lines
the corresponding results are $7.47\pm0.09, 7.48\pm0.09$,
$7.50\pm0.10$ and $7.51\pm0.10$.
As expected the stronger and partly saturated 
lines show a greater dependence on the resolution, since those
lines are sensitive to the line broadening contributed by the 
convective Doppler shifts. 
As a consequence there is a more pronounced trend between line
strength and the derived abundance with poorer resolution, as
shown in Fig. \ref{f:abu3d}, which is also reflected in an increased
scatter. Even at the highest resolution there
is still a minor trend present but it is further diminished 
when considering line profiles rather than equivalent widths 
(Paper II). A possible explanation for this may be 
that the measured Fe\,{\sc i} equivalent
widths of Blackwell et al. (1995) suffer from 
a slightly too high continuum placement. This would also partly explain
the difference between the Blackwell et al. (1995) and Holweger et al. (1995)
results for Fe\,{\sc i}, since the former sample is more biased towards
lines sensitive to the non-thermal broadening. Such a conclusion is supported by
the smaller trend for Fe\,{\sc ii} lines (Fig. \ref{f:abu3d}) 
with equivalent widths taken from Hannaford et al. (1992).
We expect that the remaining trends will vanish further at an even 
higher resolution in the simulation but there may also be an influence from
departures from LTE for the intermediate strong lines (Paper II); 
we note that no trend is apparent
for Si\,{\sc i} or Fe\,{\sc ii} lines 
at the present best resolution (Asplund 2000, Paper II).

The very strong 
Fe\,{\sc i} lines with pronounced damping wings on the other hand 
show a very
small variation between the different simulations, which follows
naturally by the robustness of the photospheric temperature and
gas pressure structure 
and thus the collisional broadening. In terms of abundances the
differences between the various simulations only amount to $\la 0.02$\,dex.
Strong lines are therefore well suited for accurate log\,$g$
determinations (e.g. Edvardsson 1988; Fuhrmann et al. 1997)
using simulations of even very modest resolution, due to the consistent
results in terms of abundances derived from weak and strong Fe\,{\sc i} lines
(Paper II).

According to common wisdom the Balmer line strengths in late-type stars
are essentially a diagnostic of the atmospheric temperature structure
(e.g. Gray 1992), since the pressure-sensitivity of the Stark-broadening
cancels the pressure-dependence on the number of H$^-$ ions in 
the ratio of line opacity to continuous opacity, leaving the necessary excitation
of H\,{\sc i} as a probe of the temperature structure. 
Given the immunity of the mean temperature to the resolution, in particular in
the relevant line-forming layers
(Fig. \ref{f:temp3d}), one would therefore expect the
wings of the Balmer lines to be independent to such details.
It does not come as a surprise then that the predicted 
H$\alpha$ and H$\beta$ lines are indeed identical in the different
simulations. Hydrogen lines can therefore with confidence be 
employed for more realistic $T_{\rm eff}$ 
determinations based on 3D simulations
even of very limited resolution, provided of course that the theoretical
(atomic) H broadening is properly understood (Barklem et al. 2000a).

\subsection{Effects on line shifts and asymmetries}
 
\begin{figure}[t]
\resizebox{\hsize}{!}{\includegraphics{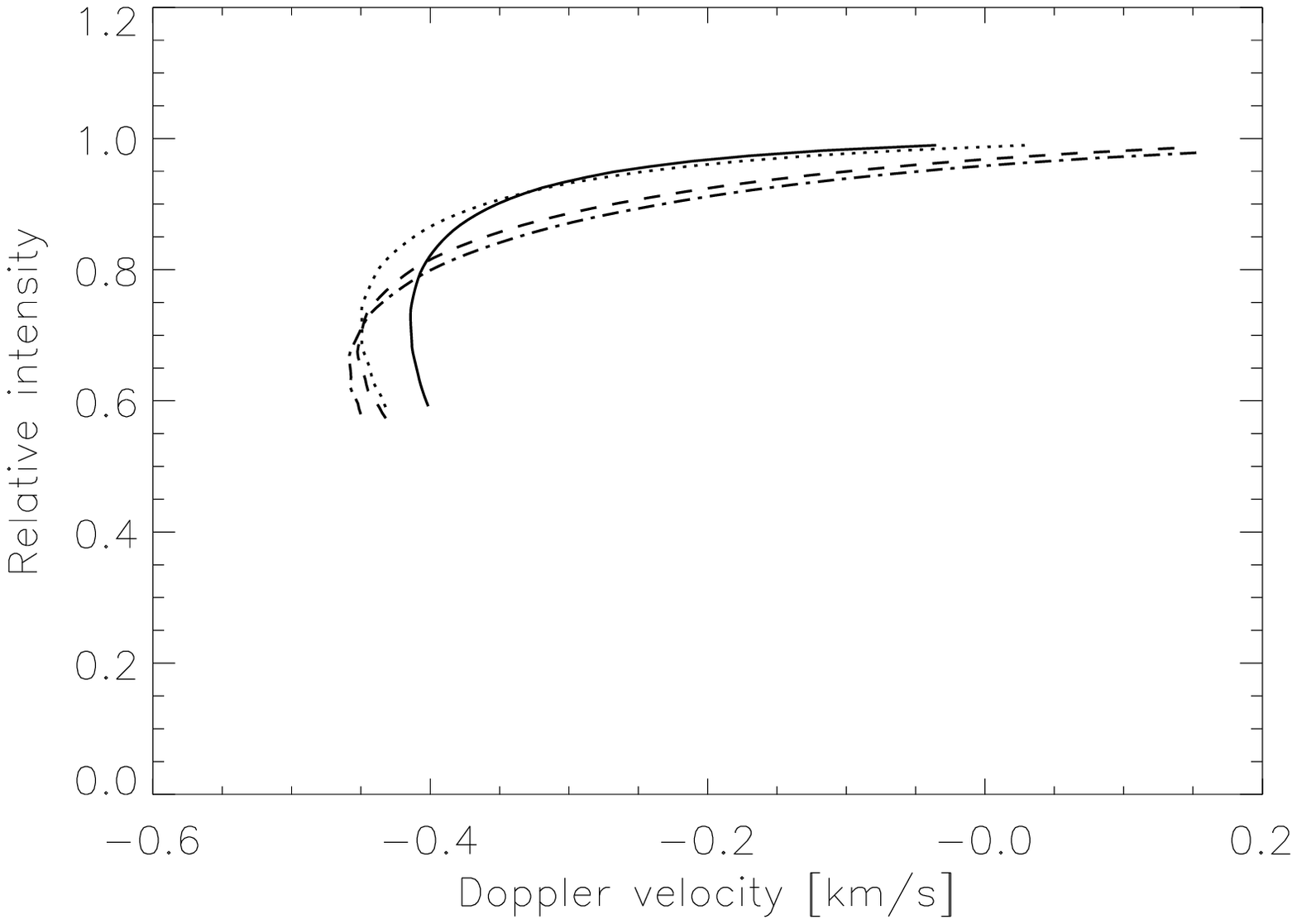}}
\resizebox{\hsize}{!}{\includegraphics{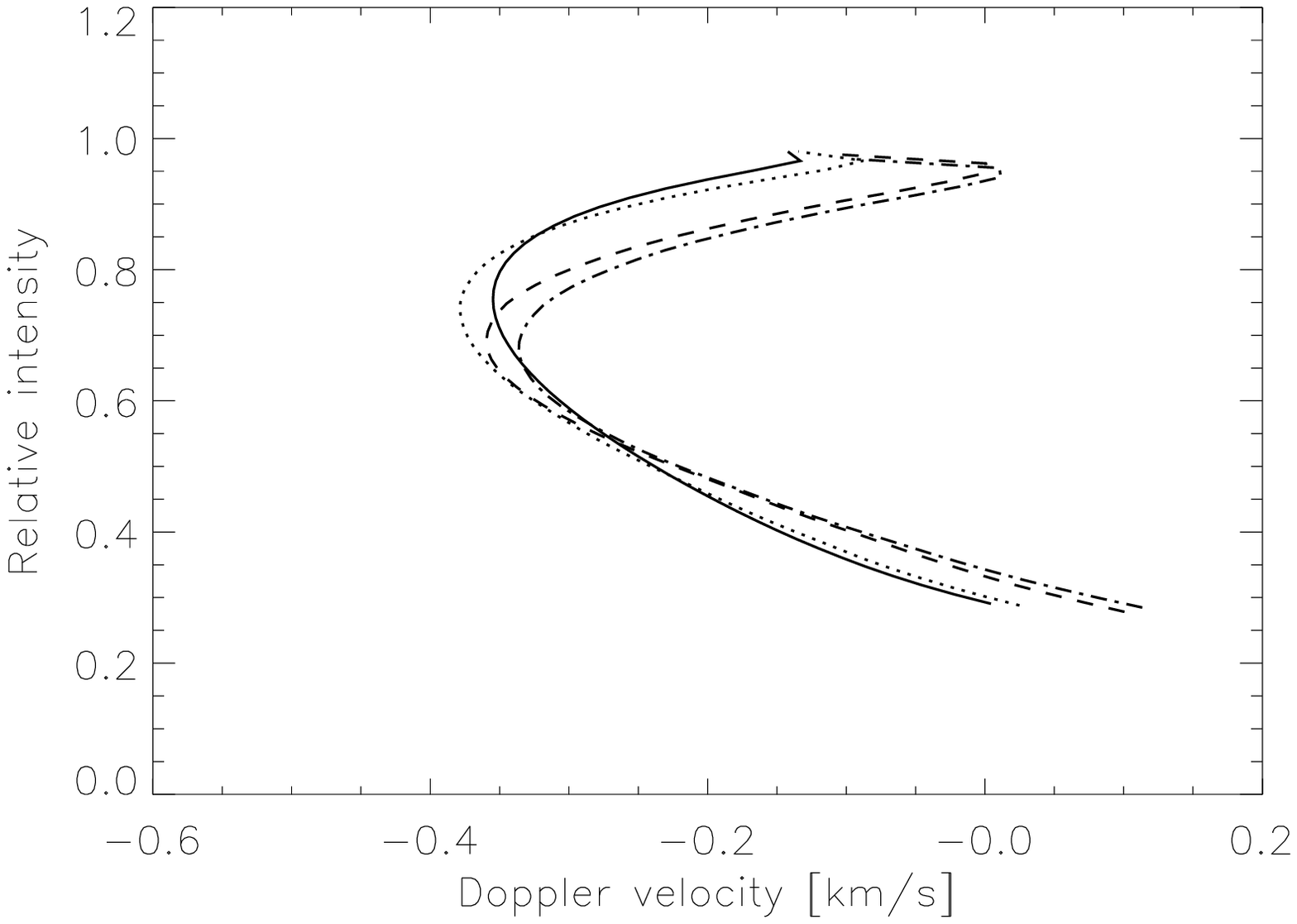}}
\caption{The predicted bisectors for the 
Fe\,{\sc i} 608.2 ({\it Upper panel}) and 
621.9\,nm ({\it Lower panel}) 
lines at different resolutions of the solar convection simulation:
200\,x\,200\,x\,82 (solid), 100\,x\,100\,x\,82 (dotted),
50\,x\,50\,x\,82 (dashed) and 50\,x\,50\,x\,63 (dot-dashed). 
All profiles have been computed with log$\,\epsilon_{\rm Fe}=7.50$. Since with
the highest resolution the theoretical bisectors agree almost perfectly 
with the observed bisectors (Paper I) it is clear that
an insufficient resolution produces discrepant line asymmetries
}
         \label{f:bis3d}
\end{figure}

The departures from perfect symmetry in the spectral line profiles
reflect the convective motion and the temperature-velocity correlations
in the line-forming region and thus
serve as sensitive probes of the detailed photospheric structure.
For the Sun, spectral line bisectors have characteristic $\subset$-shapes
although weaker lines only show the upper part of the $\subset$.
At the currently best available numerical resolution (200\,x\,200\,x\,82) 
the predicted
solar line shifts and asymmetries agree very well in general with observations
(Paper I), which therefore lends very strong support for
the realism of the solar convection simulations. 

The differences in vertical velocities (Figs. \ref{f:vz3d} and \ref{f:vzdist3d})
at different numerical resolutions manifest themselves in slight but
noticeable differences in the predicted line shifts and asymmetries,
as exemplified in Fig. \ref{f:bis3d}. Although the overall bisector shapes are
similar, the details are not. In particular the lowest resolution 
simulation produces deviant bisectors, while the intermediate cases resemble
more closely the results obtained with the highest 
resolution. It therefore seems like the simulations have nearly 
converged in terms of detailed line asymmetries. 
The differences in predicted bisectors compared to the best case
amount to $\la 200$\,m\,s$^{-1}$ for the 50\,x\,50\,x\,63 simulation.
$\la 150$\,m\,s$^{-1}$ for the 50\,x\,50\,x\,82 simulation
and  $\la 50$\,m\,s$^{-1}$ for the 100\,x\,100\,x\,82 simulation. 
In particular the disagreement is largest at line center and close
to the continuum, which also means that in order to predict
accurate stellar line shifts ($\la 50$\,m\,s$^{-1}$) 
a high numerical resolution ($\sim 100^3$)
for the 3D convection simulations is necessary. 
The remaining discrepancies in the cores of strong lines even at the
highest resolution is most likely due to the influence from the outer
boundary or possibly departures from LTE (Paper I). It should be remembered,
however, that at the heights where the cores of such strong lines are
formed the convection simulations are likely the least realistic due to
missing ingredients, such as departures from LTE, influence of magnetic fields
and additional radiative heating due to strong Doppler-shifted lines. 

\begin{figure*}[t]
\resizebox{\hsize}{!}{\includegraphics{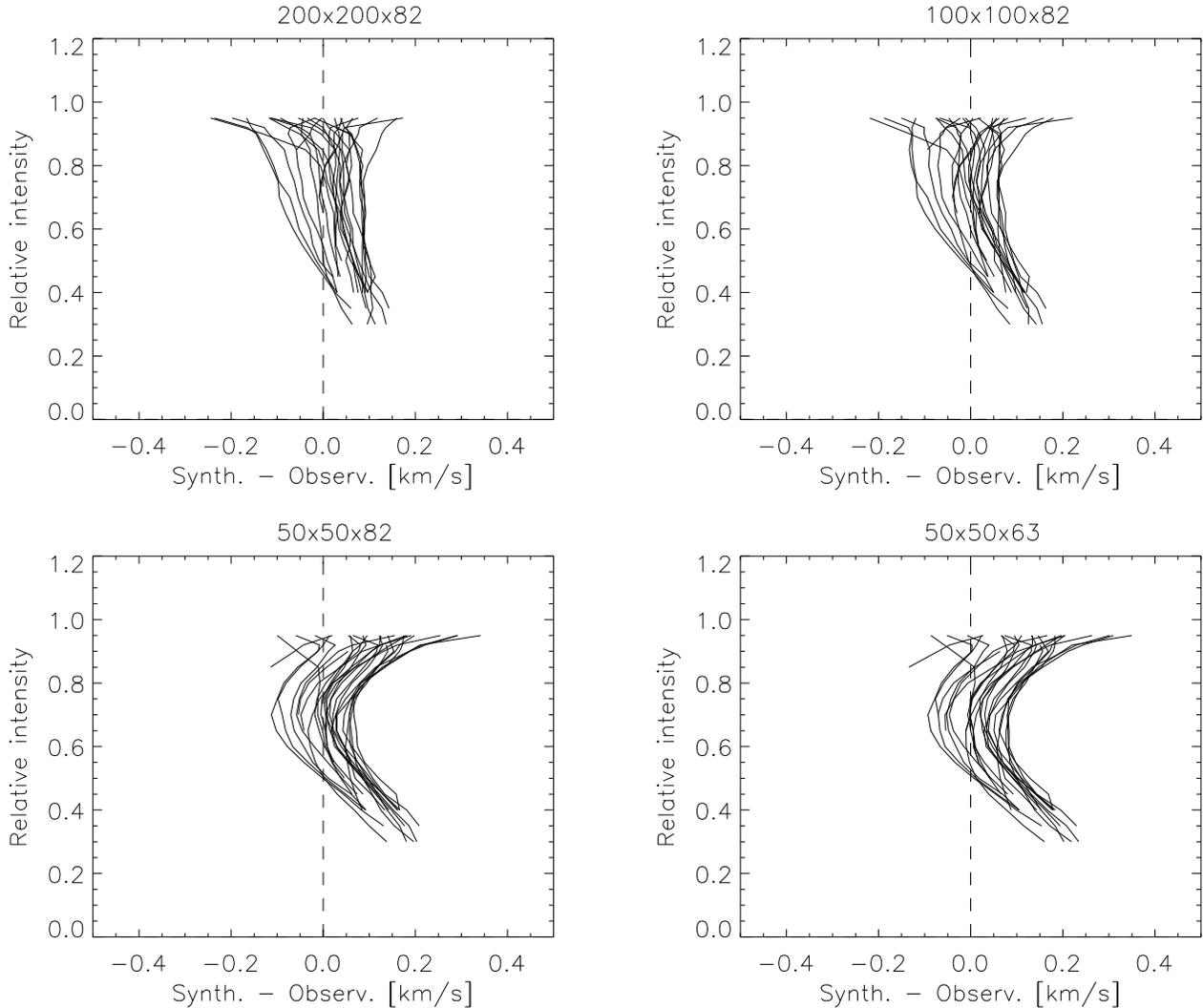}}
\caption{The {\em differences} between predicted and observed Fe\,{\sc i}
line bisectors for
different resolutions in the 3D solar convection simulations: 
{\bf a:} 200\,x\,200\,x\,82, 
{\bf b:} 100\,x\,100\,x\,82,
{\bf c:} 50\,x\,50\,x\,82, and
{\bf d:} 50\,x\,50\,x\,63
}
         \label{f:bisdiff3d}
\end{figure*}

Fig. \ref{f:bisdiff3d} shows the differences between the predicted
and observed bisectors for the Fe\,{\sc i} lines in Table \ref{t:lines}
for the various 3D resolutions. Systematic 
shortcomings are apparent for the poorer resolution but the
situation gradually improves with higher resolutions. In particular
for the most extensive simulation, most predicted bisectors agree
very nicely with the observed line asymmetries (the increased scatter
close to the continuum is due to the influence of weak blends). 
The inclusion of additional Fe\,{\sc i} lines corroborate 
this conclusion (Paper I).
Clearly both a high vertical and horizontal resolution is necessary
to have accurate predictions in terms of line asymmetries. 

\section{Comparison between 2D and 3D}

\subsection{Effects on continuum intensity contrast \label{s:irms2d}}
 
The restriction to 2D forces the over-turning motion to take place
only in one horizontal direction rather than two, which in turn
produces different typical length scales and convective velocities
in the photosphere compared to 3D. According to Fig.~\ref{f:intpower}
the granular scales are slightly larger in 2D as a result of the
larger horizontal pressure fluctuations (Ludwig et al., in preparation),
which also implies that the intensity contrast will be different
relative to the 3D case in order to maintain the larger spatial scales. 
The 2D, $I_{\rm rms}$ is notably 
larger than for the corresponding 3D simulation even with the
same effective viscosity (i.e. the same horizontal resolution dx): 
17.5\% and 16.6\%, respectively. 
This difference partly explains the discrepant line asymmetries in 2D,
as shown in Sect. \ref{s:asym2d}.

\subsection{Effects on line shapes and abundances \label{s:shape2d}}

\begin{figure}[t]
\resizebox{\hsize}{!}{\includegraphics{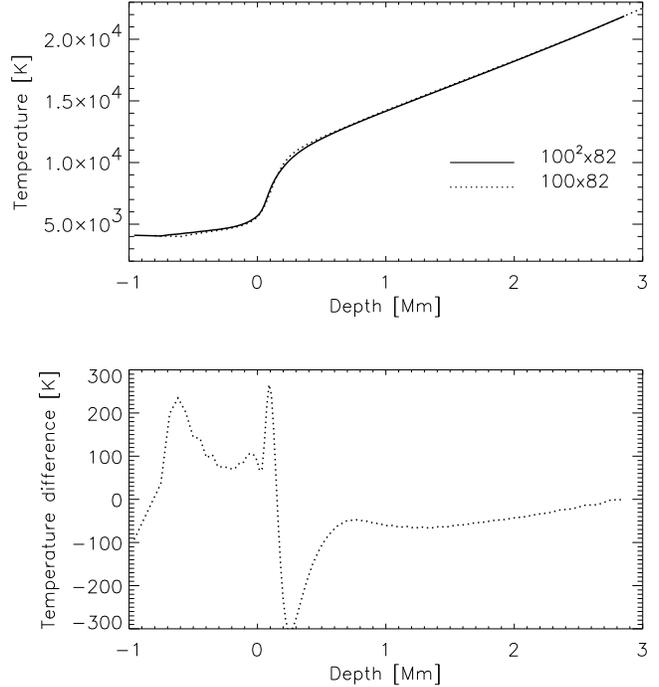}}
\caption{{\it Upper panel:}
The mean temperature structure in solar convection simulations of 
different number of dimensions:  3D (solid) and 2D (dotted).
{\it Lower panel:} The temperature difference in the sense 3D-2D,
i.e. the temperatures tend to be higher in 2D in the deeper layers and
lower in the optically thin regions
}
         \label{f:temp2d}
\end{figure}

\begin{figure}[t]
\resizebox{\hsize}{!}{\includegraphics{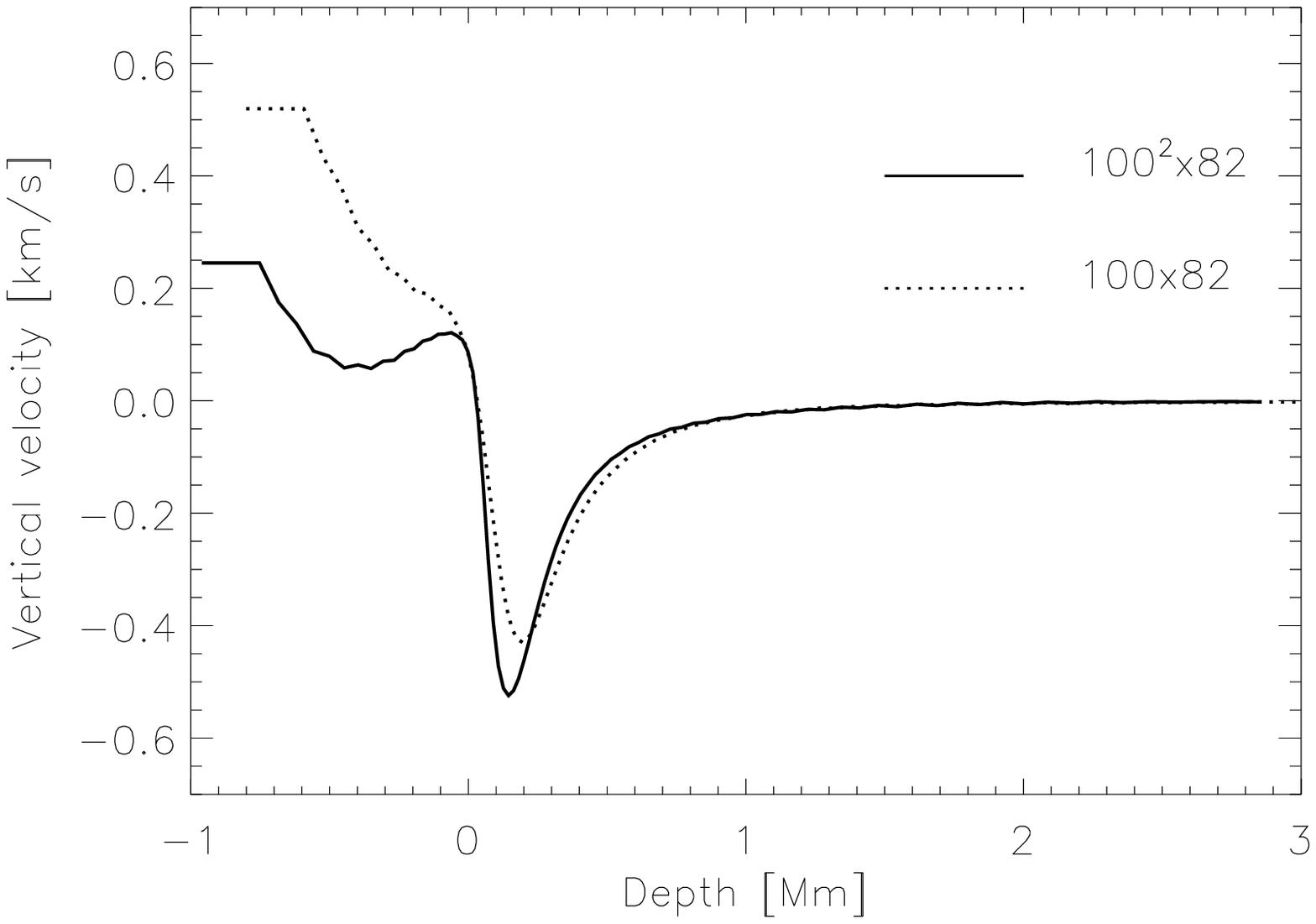}}
\resizebox{\hsize}{!}{\includegraphics{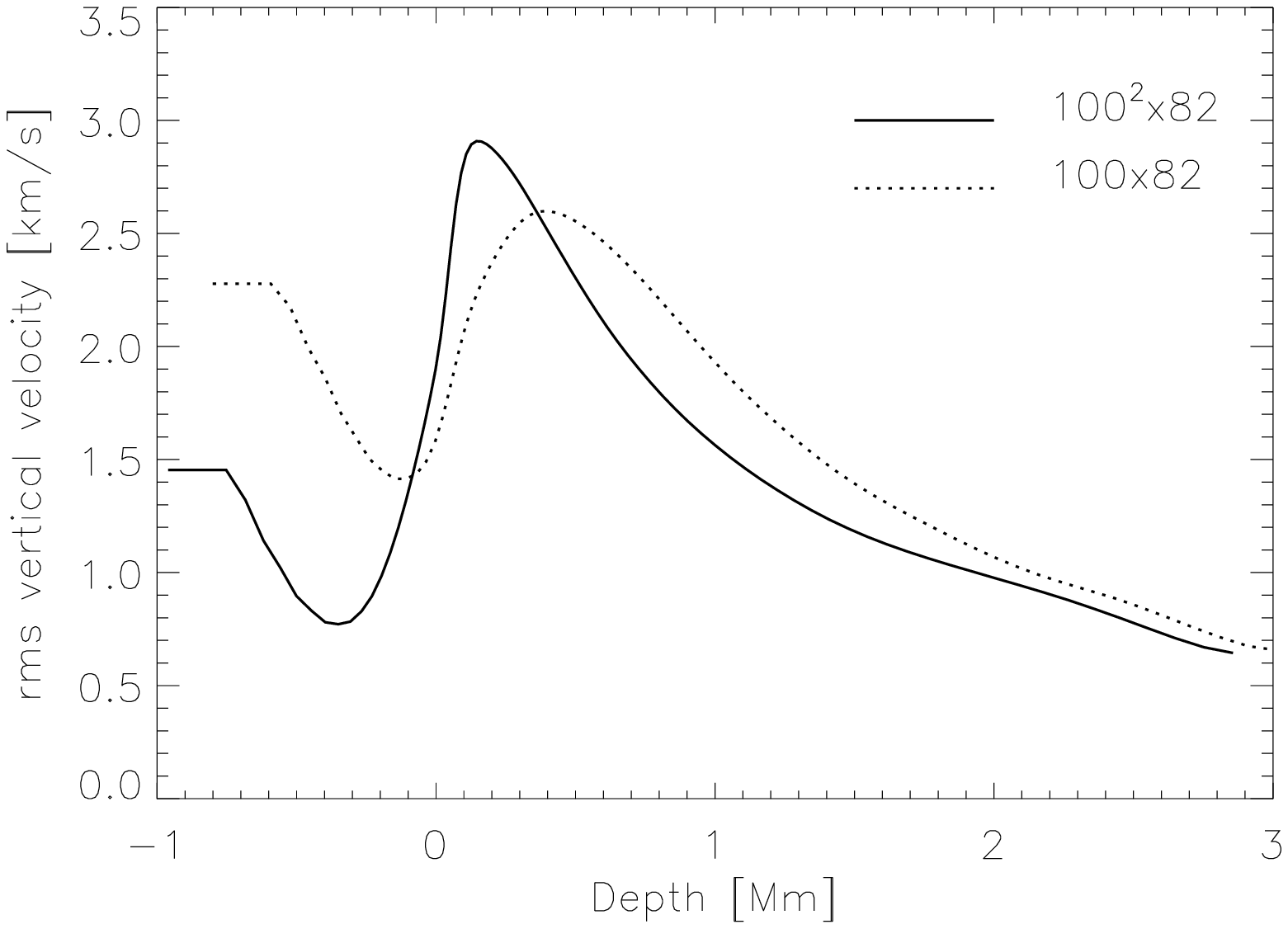}}
\caption{The time-averaged mean vertical velocity ({\it upper panel}) and 
horizontal rms vertical velocity ({\it lower panel}) 
in solar convection simulations of different number of dimensions:  
3D (solid) and 2D (dotted).
Positive vertical velocities correspond to downflows
}
         \label{f:vz2d}
\end{figure}

\begin{figure}[t]
\resizebox{\hsize}{!}{\includegraphics{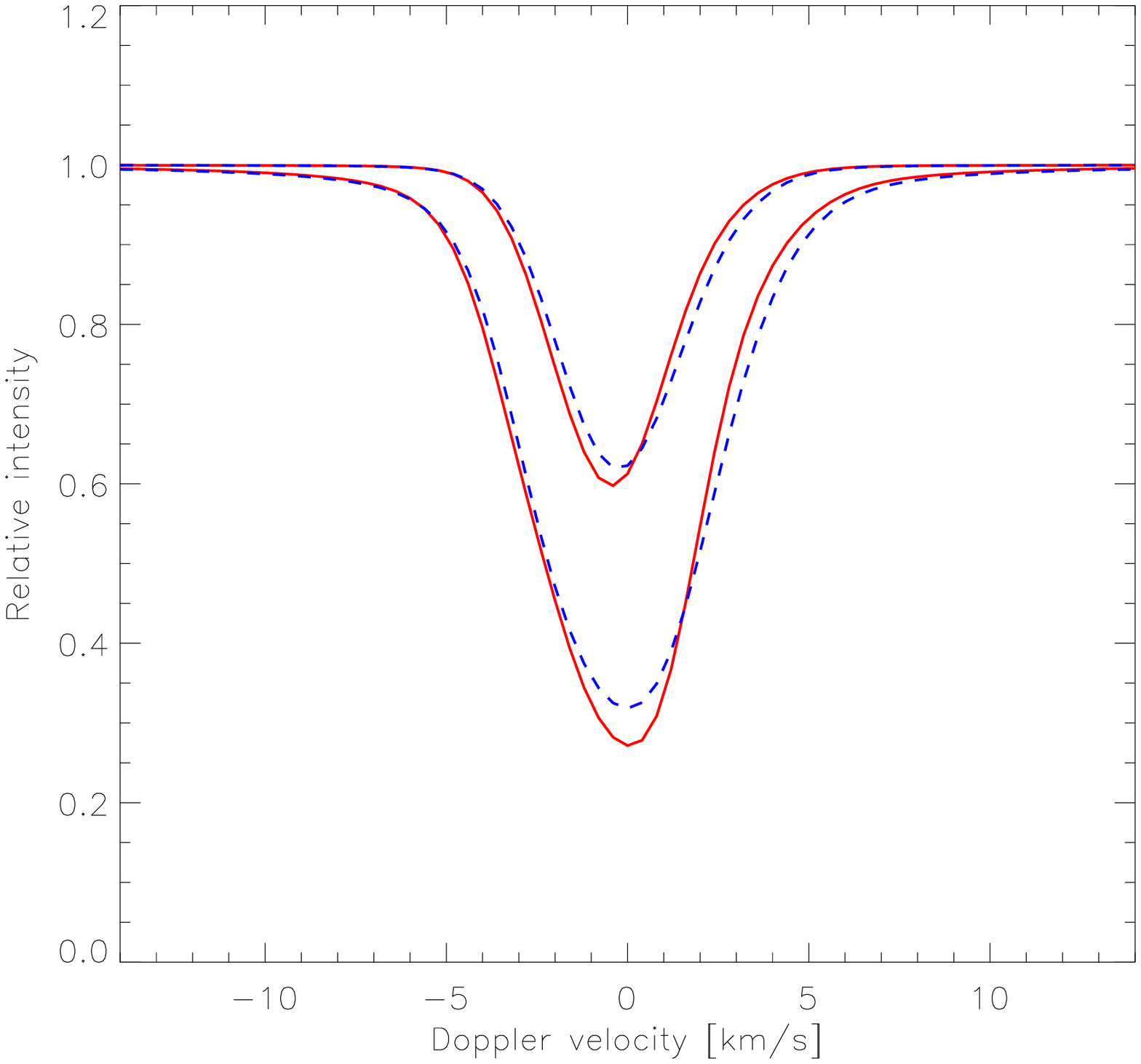}}
\caption{The predicted Fe\,{\sc i} 608.2 (weaker) and 621.9\,nm (stronger) 
lines at different number of dimensions of the solar convection simulation:
3D (solid) and 2D (dashed). 
The individual Fe abundances have been adjusted according to Table \ref{t:lines}
to produce the same equivalent widths in 2D and 3D
}
         \label{f:prof2d}
\end{figure}

Due to the smaller computational demands posed by 2D convection simulations
and radiative transfer computations
compared to corresponding 3D calculations, it is of interest to
compare the effects of number of dimensions on the predicted spectral
line profiles. 
Since the temperature and velocity structures depend on 
the number of dimensions as seen in Figs. \ref{f:temp2d} and \ref{f:vz2d},
it can not be assumed that the predicted line shapes and strengths 
will be the same in 2D and 3D.
The restriction to 2D facilitates a more efficient merging of down-flows. 
This produces larger horizontal scales in 2D (Fig. \ref{f:intpower}, 
which is associated with larger horizontal pressure fluctuations
and thus different vertical velocity structures. The details of the
differences in
the temperature structures are more subtle but are related to the 
velocity variations; 
cf. Ludwig et al., in preparation, for a detailed discussion on the
physical reasons for the differences in convection properties between
2D and 3D.
 
Fig. \ref{f:prof2d} shows a couple of theoretical 2D and 3D  Fe\,{\sc i} lines 
computed with the Fe abundances given in Table \ref{t:lines} to return
the same equivalent widths; the Fe\,{\sc ii} lines show
the same behaviour and discrepancies. 
Although the overall line shape (but not the detailed line 
asymmetries and shifts as discussed in 
Sect. \ref{s:asym2d}) is relatively similar 
for weaker Fe\,{\sc i} lines, the profiles of intermediate strong 
lines are very
different, with the 2D profiles being much shallower and broader than
the corresponding 3D profiles. Since in 3D the predicted lines agree
almost perfectly with observations (Paper I and Paper II), 
it implies that the agreement is
relatively poor in 2D. In particular, even at very high resolution in
2D abundance analyses would have to be restricted to using equivalent widths
rather than profile fitting due to the inherent shortcomings of the 2D predictions.
Likewise, with 2D analyses too low projected rotational velocities of the stars
would be obtained due to the too broad predicted line profiles.

\begin{figure*}[t]
\resizebox{\hsize}{!}{\includegraphics{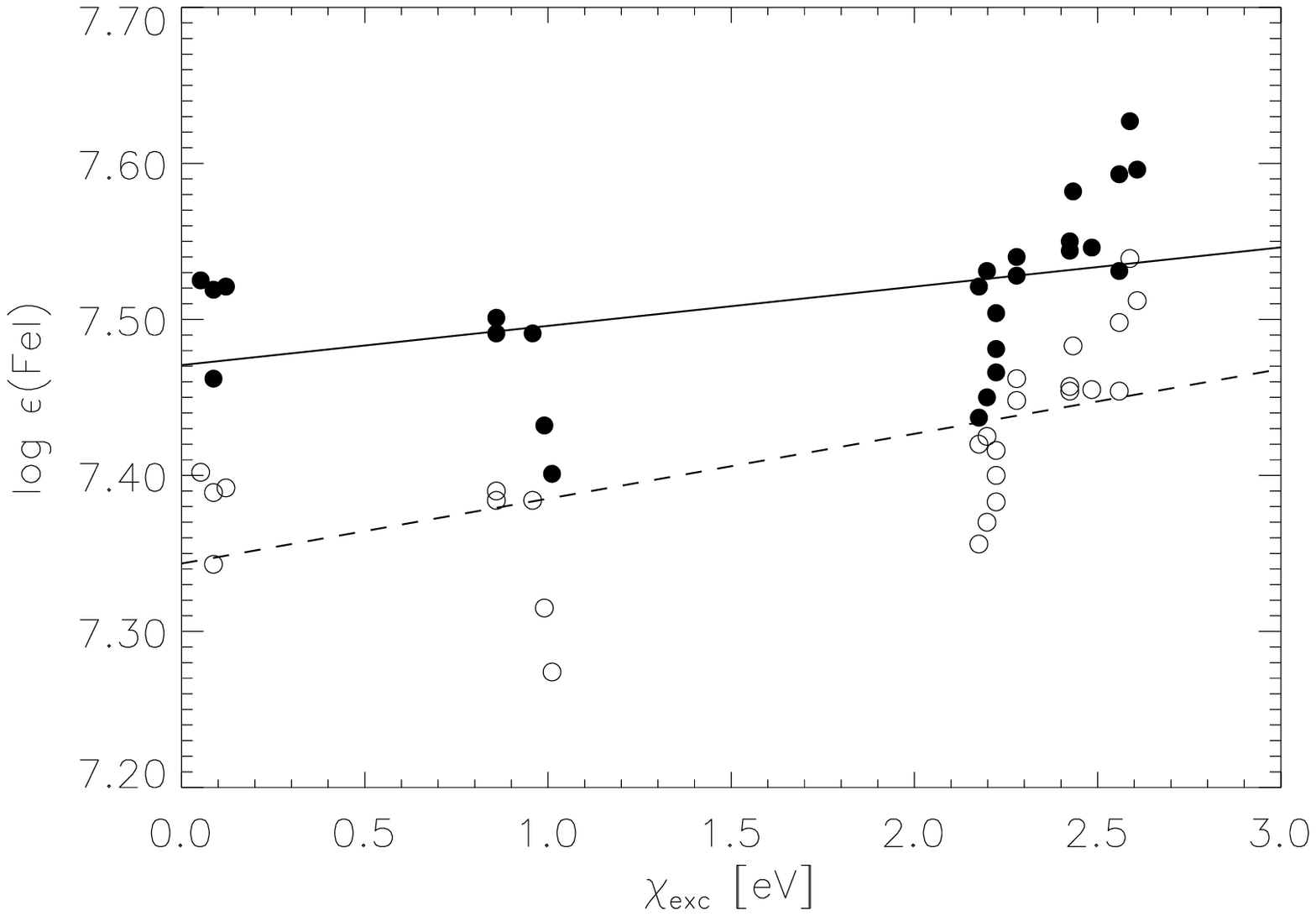}
                      \includegraphics{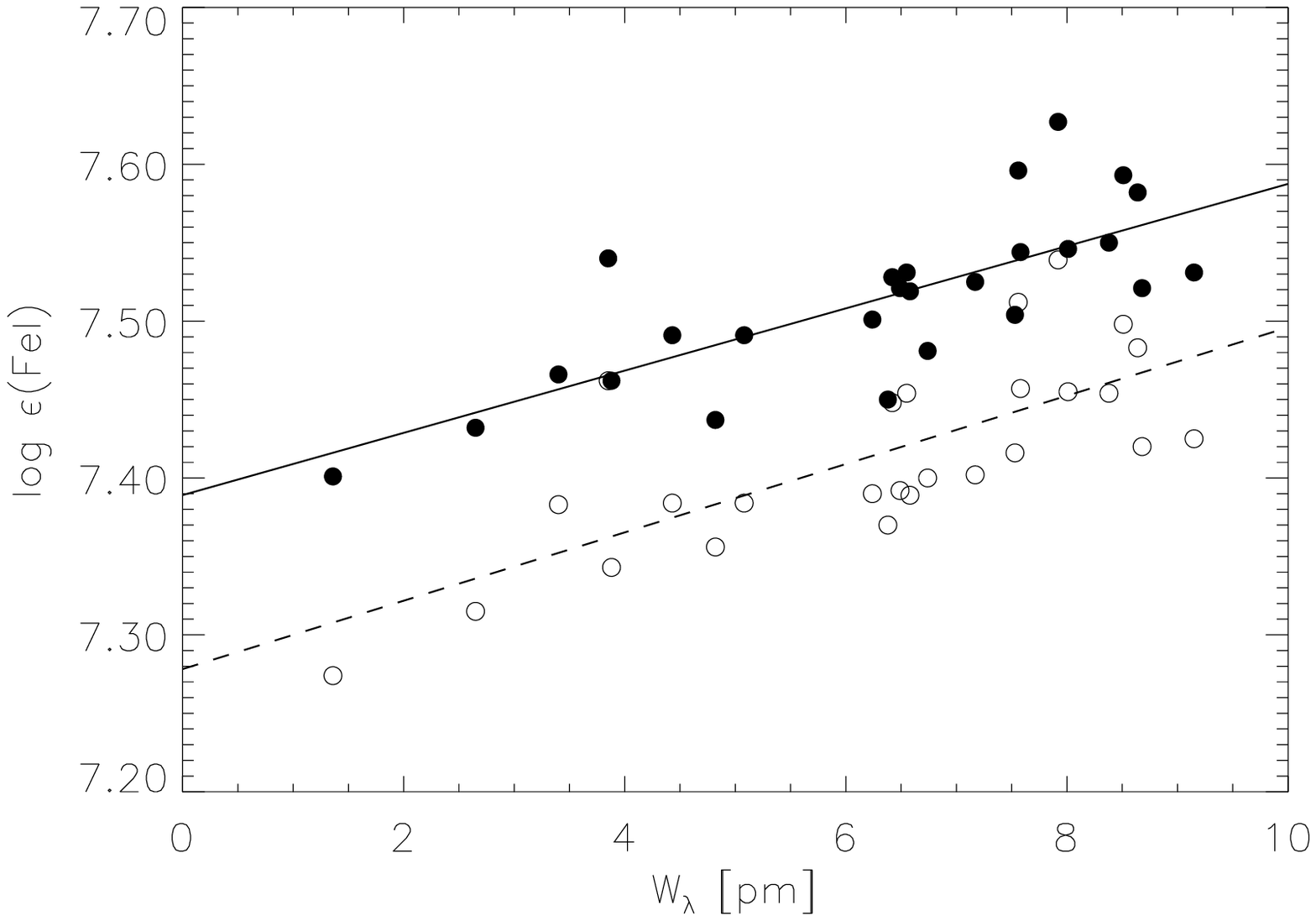}}
\resizebox{\hsize}{!}{\includegraphics{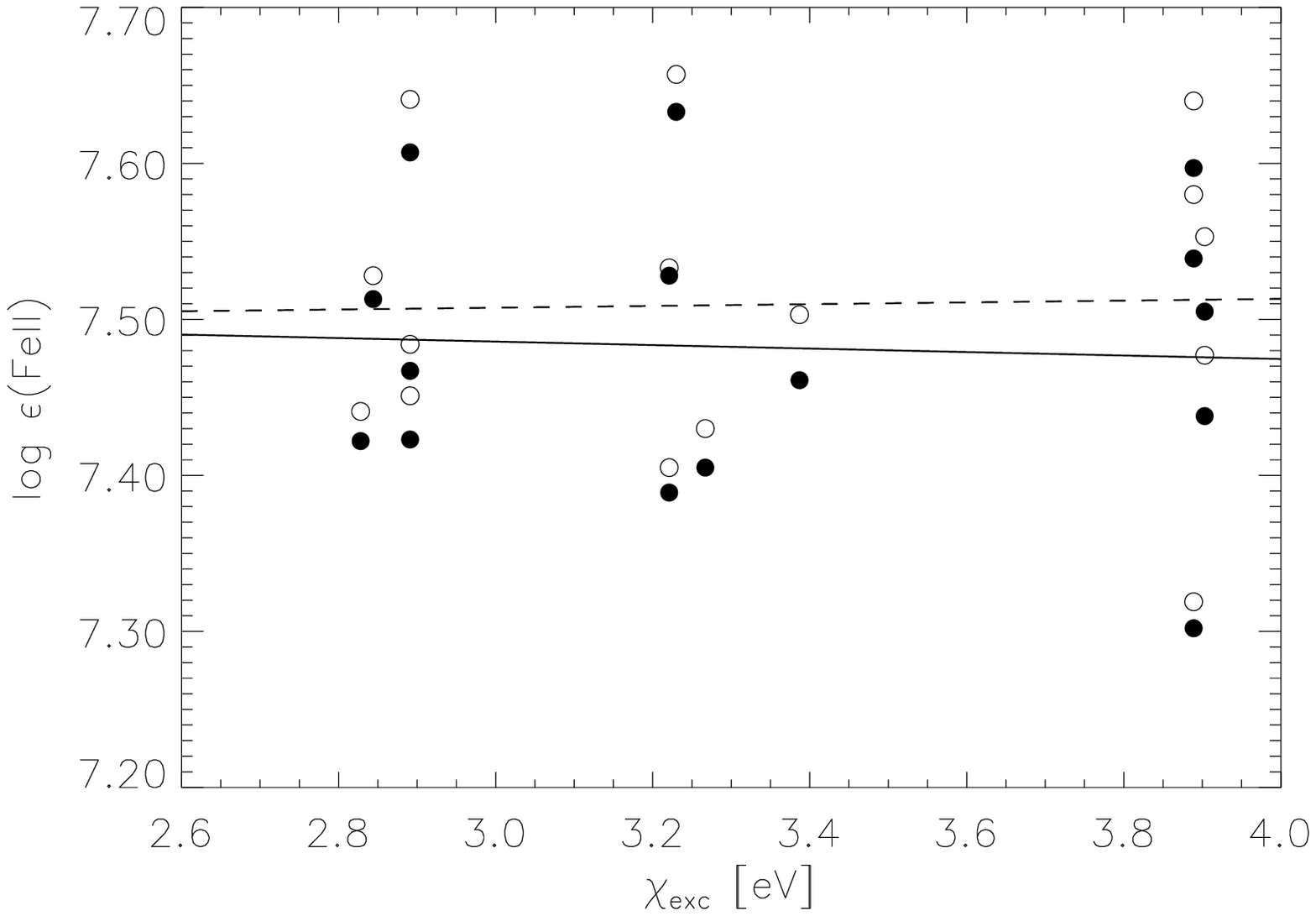}
                      \includegraphics{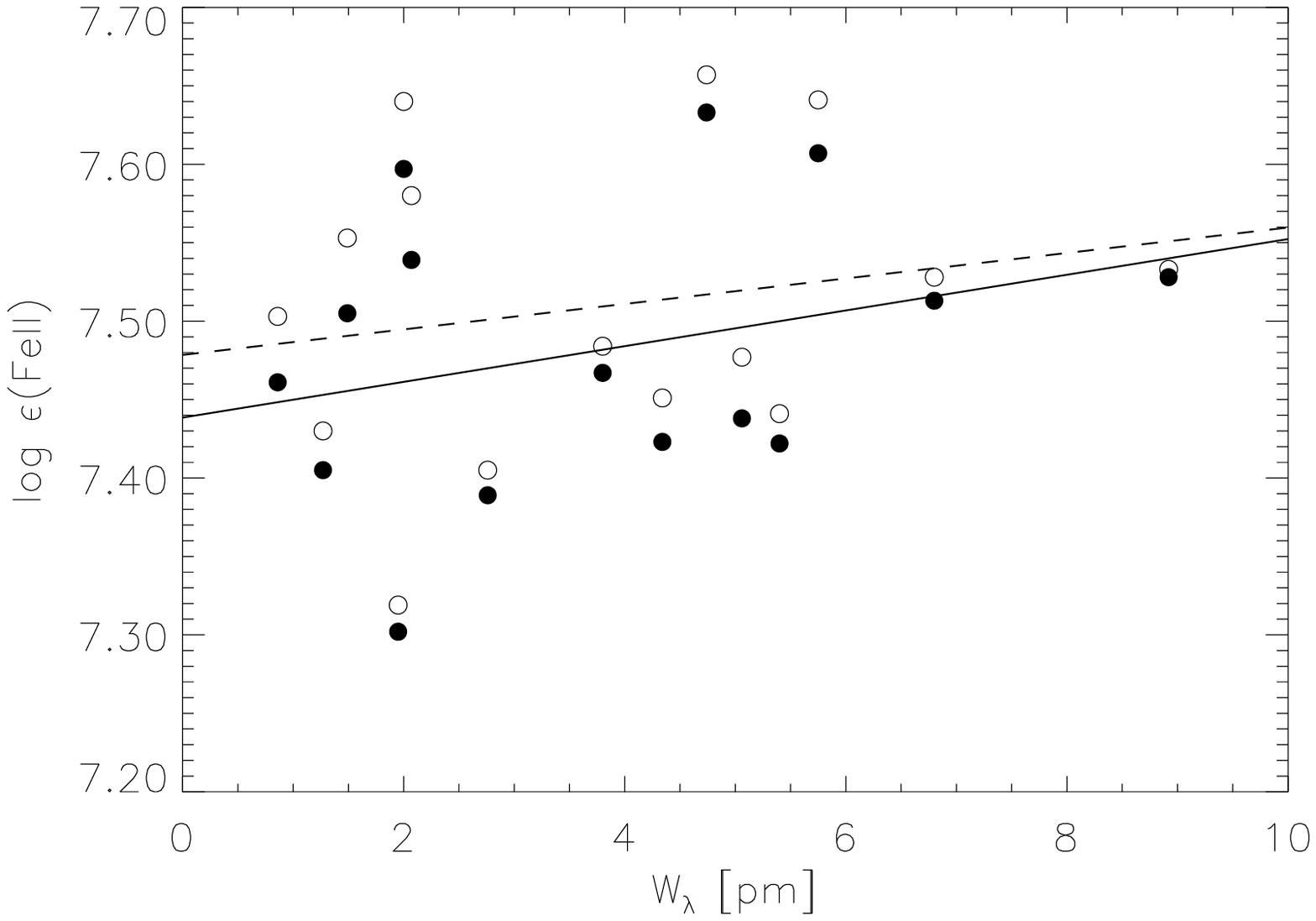}}
\caption{The abundances derived from Fe\,{\sc i} ({\it upper panel}) and 
Fe\,{\sc ii} lines ({\it lower panel}) as a function of the
excitation potential ({\it left panel}) and equivalent widths 
({\it right panel}) 
for different number of dimensions of the convection simulations:
3D (filled circles) and
2D (open circles). 
The solid and dashed lines are least-square fits
to the two sets of abundances. The trend with line strength is more pronounced
here than in Paper II due to the use of equivalent widths instead of profile
fitting and lower numerical resolution
}
         \label{f:abu2d}
\end{figure*}

In terms of derived abundances the 2D and 3D results show systematic differences.
The sample of Fe\,{\sc i} lines
in Table \ref{t:lines} give mean Fe abundances of 
$7.51\pm0.05$ (3D: 100\,x\,100\,x\,82) and $7.41\pm0.06$ (2D: 100\,x\,82). 
The corresponding results for the Fe\,{\sc ii} lines are
$7.48\pm0.09$ (3D) and $7.51\pm0.10$ (2D).
The remaining difference in mean $T_{\rm eff}$ of 36\,K between the 2D and 3D case 
translates to an abundance difference for Fe\,{\sc i} and Fe\,{\sc ii} of
-0.03 and +0.01\,dex, respectively.
The larger difference between the 2D and 3D cases for Fe\,{\sc i} lines 
is a natural
consequence of the greater sensitivity to the temperature structure for those lines.
The derived 2D Fe abundances show essentially identical 
trends with equivalent widths and excitation potential, as seen in
Fig. \ref{f:abu2d}; the differences in the Fe\,{\sc i} results are
basically restricted to a systematic offset of about 0.1\,dex for all lines. 
With the current best solar simulation (the 200\,x\,200\,x\,82
simulation used here) and profile fitting instead of equivalent widths, neither
the Fe\,{\sc i} nor the Fe\,{\sc ii} lines show any significant trends with
excitation potential, and only Fe\,{\sc i} lines depend slightly
on line strengths, which may reflect departures from LTE
rather than shortcomings of the convection simulations as such (Paper II). 

Strong Fe\,{\sc i} lines with pronounced pressure damped wings can
be efficient gravitometers (e.g. Edvardsson 1988; Fuhrmann et al. 1997),
provided the collisional broadening is properly understood 
(Anstee \& O'Mara 1991, 1995; Barklem \& O'Mara 1997;
Barklem et al. 1998, 2000b). Due to differences in pressure and
temperature structures, the predicted strong lines are indeed quite
different in 2D compared to in 3D, with the 2D lines being
stronger for a given abundance. While in 3D the strong Fe\,{\sc i} lines
imply abundances consistent within 0.05\,dex of those of weak
and intermediate strong lines (Paper II), in 2D the strong lines suggest abundances 
systematically about 0.10\,dex lower than those of weaker lines.
This naturally translates to
significant errors when estimating stellar surface gravities 
from the wings of strong lines using 2D convection simulations.

\begin{figure}[t]
\resizebox{\hsize}{!}{\includegraphics{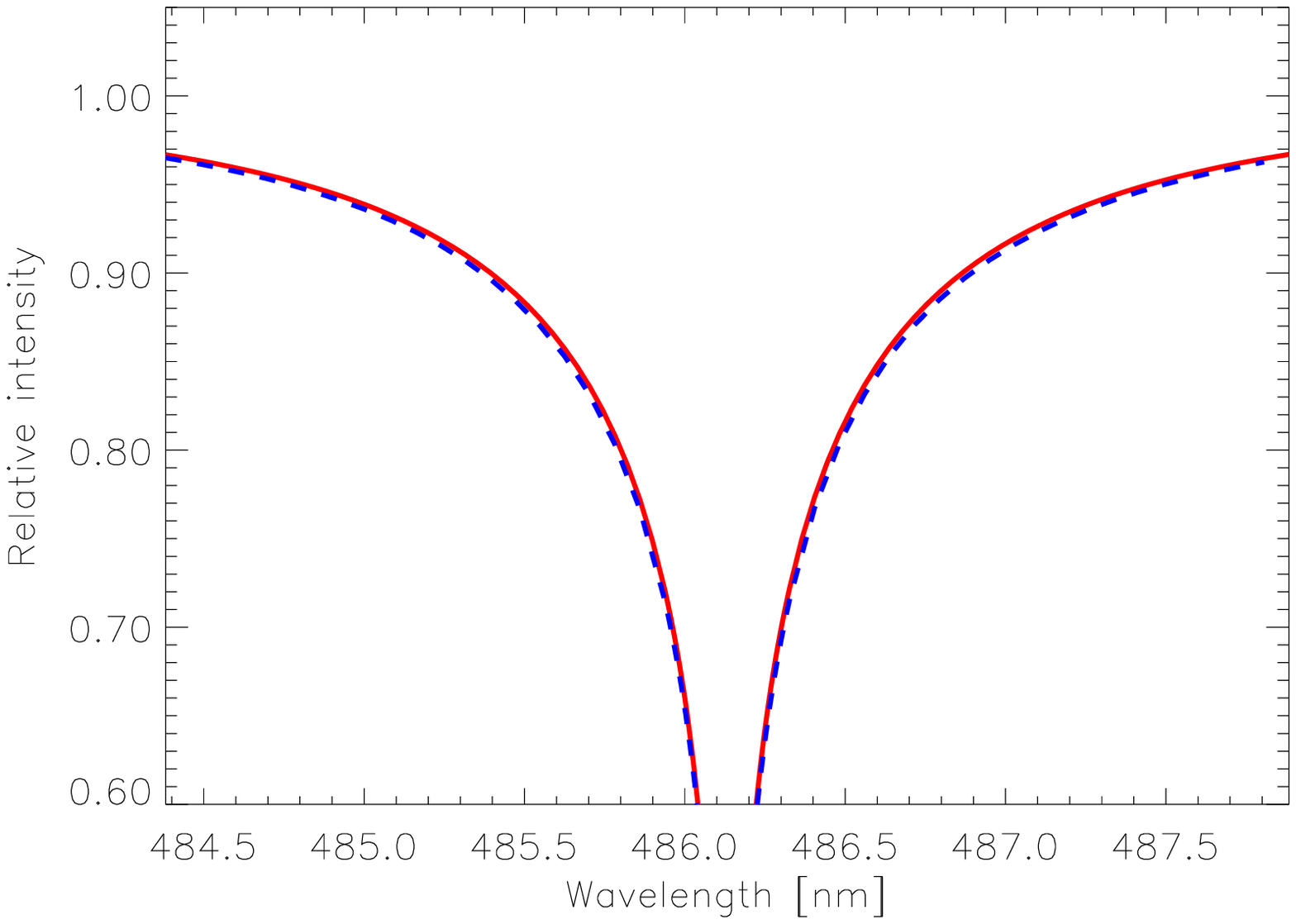}}
\caption{The temporally and spatially averaged H$\beta$ profile computed
using the 3D (solid line) and 2D (dashed line) solar convection
simulations. In spite of the existing temperature differences 
(Fig. \ref{f:temp2d}), the spatially averaged H$\beta$ profiles are very
similar in 2D and 3D
}
         \label{f:hbeta2d}
\end{figure}

In 3D the H lines are effectively independent on the chosen
resolution as discussed above. When comparing 2D with 3D the
same conclusion does not follow automatically since the convective 
transport properties and therefore temperature structures 
are different, as illustrated in Fig. \ref{f:temp2d}
and discussed in detail by Ludwig et al., in preparation. The temperature 
difference is notably larger between 2D and 3D than for different numerical
resolution in 3D (Fig. \ref{f:temp3d}).
As seen in Fig. \ref{f:hbeta2d} the resulting Balmer H$\beta$ profiles
are however very similar, albeit with minor differences in the near wing,
with the 2D profile being slightly broader than in 3D, in accordance with the
somewhat steeper temperature gradient (Fig. \ref{f:temp2d}). 
Since in terms of $T_{\rm eff}$ the difference only amounts to about 30\,K, 
an accurate $T_{\rm eff}$ calibration using Balmer
lines should be possible
with 2D convection simulations, provided the remaining problems with
the theoretical (atomic) 
H line broadening can be addressed (Barklem et al. 2000a).

\subsection{Effects on line shifts and asymmetries
\label{s:asym2d}}
 
\begin{figure}[t]
\resizebox{\hsize}{!}{\includegraphics{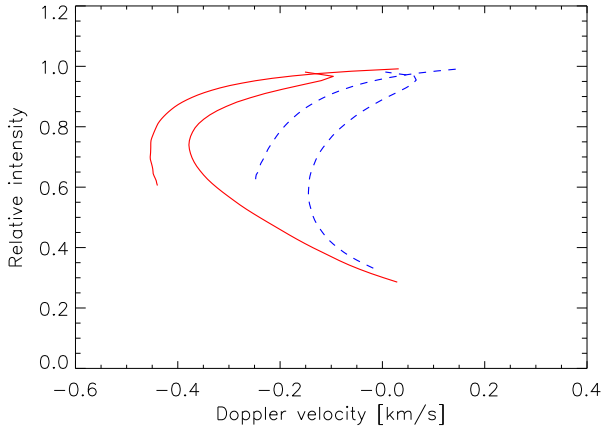}}
\caption{The predicted bisectors for the 
Fe\,{\sc i} 608.2 (weaker) and 
621.9\,nm (stronger) 
lines at different number of dimensions of the solar convection simulation:
3D (solid) and 2D (dashed). 
The individual Fe abundances have been adjusted according to Table \ref{t:lines}
to produce the same equivalent widths in 2D and 3D.
Since in 3D the theoretical bisectors agree almost perfectly 
with the observed bisectors (Paper I) it is clear that
2D simulations produce discrepant line asymmetries. Note that all 
bisectors are on an absolute wavelength scale, emphasizing the shortcomings
of the 2D simulations to accurately predict the line shifts
}
         \label{f:bis2d}
\end{figure}

The atmospheric temperature inhomogeneities and macroscopic velocity
fields produce characteristic spectral line asymmetries, which 
for solar-type stars take on $\subset$-shaped bisectors.
Since both the temperatures and flow pattern are different in
2D than in 3D (Figs. \ref{f:temp2d} and \ref{f:vz2d}), the predicted bisectors
will depend on the adopted number of dimensions of the convection
simulations, which is illustrated in Fig. \ref{f:bis2d} for
a couple of Fe\,{\sc i} lines.
Although the overall bisector shapes are qualitatively similar the detailed
line asymmetries are not, neither for weak nor intermediate strong
Fe\,{\sc i} and Fe\,{\sc ii} lines. 
The predicted line shifts from the 2D simulation 
are less blue-shifted by typically 100-200\,m\,s$^{-1}$ compared
with the corresponding 3D estimates for weak lines, although the differences
vanish for stronger lines with cores formed above the granulation layers. 
Due to the complex interplay between temperature inhomogeneities,
velocity fields and brightness contrasts in producing the line asymmetries,
it is very difficult to identify which difference in the physical variables
is most responsible for the bisector variations, but it is clear that
at least the quite different convective velocities play an important role,
since the deceleration zone occurs over a larger vertical extent with
different amplitude of the vertical velocities, as seen in Fig. \ref{f:vz2d}.

\begin{figure}[t]
\resizebox{\hsize}{!}{\includegraphics{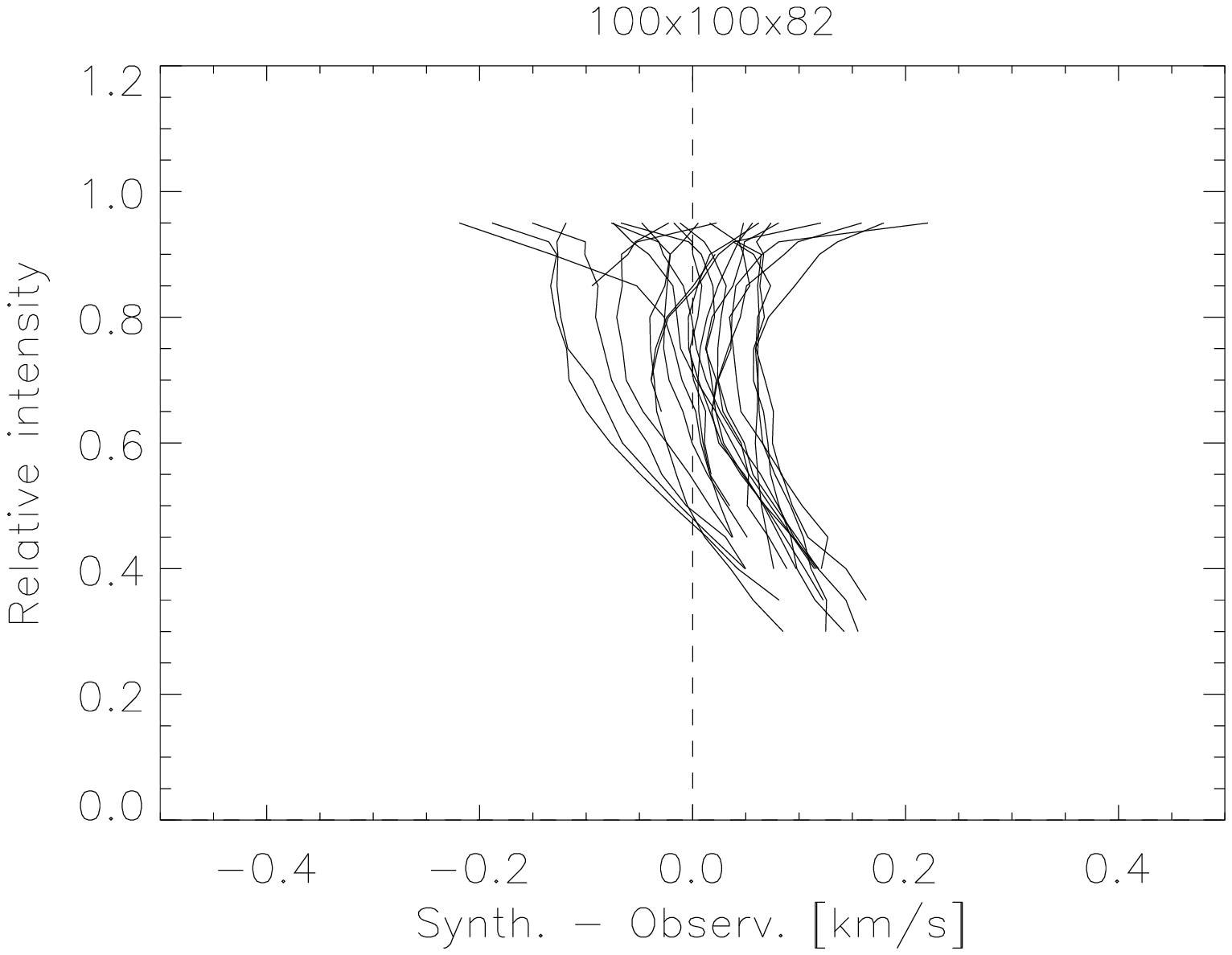}}
\resizebox{\hsize}{!}{\includegraphics{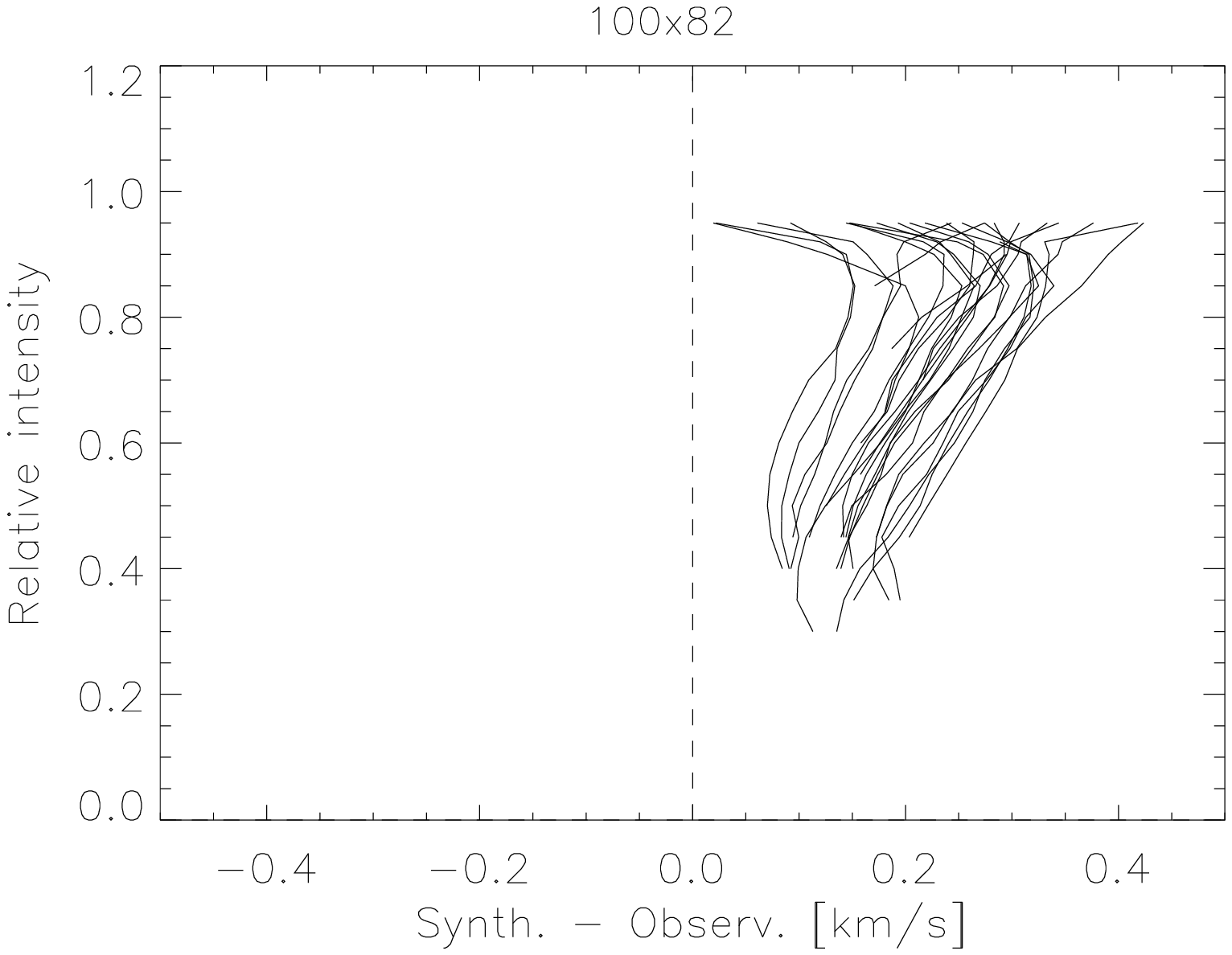}}
\caption{The {\em differences} between predicted and observed Fe\,{\sc i}
line bisectors for 3D ({\it Upper panel}) and
2D ({\it Lower panel}) solar convection simulations.
In 2D neither the line asymmetries nor the line shifts are accurately
described
}
         \label{f:bisdiff2d}
\end{figure}

The discrepancies are even larger closer to the continuum, amounting
to as much as 300\,m\,s$^{-1}$, as evident in Fig. \ref{f:bisdiff2d} which
shows the differences between theoretical and
observed solar Fe\,{\sc i} line bisectors for the corresponding 2D and 3D
simulations. It should be noted that
the time coverages are sufficient for both the
2D and 3D simulations (16.5\,hrs and 50\,min, respectively) 
to produce statistically significant spatially 
and temporally averaged bisectors, as verified by test calculations
restricted to much shorter time sequences: half the time interval
produces indistinguishable bisectors from the full calculations, while
shorter sequences covering 10\% of the whole simulation 
stretches have an accuracy of $\la 100$\,m\,s$^{-1}$ due to the influences
of granular evolution and the radial oscillations corresponding to the
solar 5-min oscillations which are present in the numerical box.
Furthermore, the blue-most bend in the bisectors also occur at
significantly larger line-depths in 2D than in 3D, again reflecting
the shortcomings when simulating a 3D phenomenon like convection in 2D.
Since the most realistic 
3D predictions agree very well with the observed bisectors
it is clear that 2D results are significantly less accurate.

\section{Conclusions}

The aim of the present paper has been to investigate how sensitive
predicted line profiles and bisectors are to the adopted numerical
resolution and number of dimensions of the convection simulations used
as model atmospheres in the line transfer calculations. 
The investigations have been performed strictly differentially in
order to isolate the effects of the dimensions of the numerical grid.

The numerical resolution in 3D has a limited influence on the theoretical
line profiles and asymmetries. With a too coarse resolution the lines
tend to be slightly 
too narrow and deep, but at the highest resolution we have used
to date the predictions have converged almost perfectly to the observed
values, both in terms of line shapes and asymmetries, which lend strong
support to the realism of the convection simulations. In terms of
abundances, weak lines show a small dependence on resolution 
($\simeq 0.02$\,dex) while intermediate strong lines with their larger
sensitivity to the non-thermal Doppler broadening show a greater
dependence ($\la 0.10$\,dex).
Both for the intention of deriving accurate elemental abundances and
using line asymmetries as probes of stellar surface convection,
a resolution of $\simeq 100^3$ appears sufficient considering the
observational uncertainties in stellar spectroscopy today
($\la 0.05$\,dex and $\la 100$\,m\,s$^{-1}$, respectively).
On the other hand, strong lines of Fe\,{\sc i}
and H are only marginally affected by the resolution, since they mainly
reflect the atmospheric temperature and pressure structures which
converge already at very modest resolution, as a natural consequence
of mass conservation (Stein \& Nordlund 1998). Therefore, accurate
$T_{\rm eff}$ and log\,$g$ calibrations can be achieved even with a grid
of 3D convection simulations of very limited resolution. 

Unfortunately 2D convection simulations appear less reliable for
abundance analyses and studies of line asymmetries, since
the inherent convective transport properties are different in 2D
compared to in 3D. The predicted line profiles in 2D are too shallow
and broad for a given line strength, in particular for intermediate
strong lines which are sensitive to the convective velocity broadening. 
As a consequence, the agreement with observed profiles is far from
satisfactory. Furthermore, the derived abundances are not immune either
to the adopted number of dimensions of the convection simulation, in particular
for the Fe\,{\sc i} lines.  
The same conclusion holds when comparing line asymmetries
and shifts with differences amounting to $\ga 200$\,m\,s$^{-1}$ on 
an absolute velocity scale. 
Even the coarsest 3D resolution investigated here (50\,x\,50\,x\,63)
produces more realistic results in general than corresponding
2D simulations in terms of line strengths and asymmetries.
In light of the findings presented here and in Paper I, one may conclude
that some of the claims of a good correspondence between observations and
2D predictions (e.g. Gadun et al. 1999) are probably fortuitous. 
The right results can be obtained
for the wrong reasons if shortcomings in terms of e.g. resolution,
equation-of-state, opacities, depth scale, 
and temperature structure compensate
the errors introduced by the restriction to 2D rather than 
treating convection fully in 3D.

Also for an additional reason, 2D convection simulations provide 
less of an attractive approach compared to 3D than at first glance
when considering the computational demand. Due to the poorer
spatial coverage of the surface convection, the time variations are
significantly larger in 2D than in 3D. As a consequence, correspondingly
longer time sequences are needed in order to obtain statistically
significant averaged convection properties and line profiles. 
Rather than a difference of a factor of about 100 (assuming typical
numerical grids of $100^3$ and $100^2$, respectively,
mesh points) in computing
time, in practice only a factor of $\la 5$ is more likely saved. 
Given the limitations found here in terms of spectral line formation,
the difference is hardly of great importance, in particular since
more realistic 3D simulations are affordable today even with 
normal work-stations.

\begin{acknowledgements}
Illuminating discussions with B. Freytag and B. Gustafsson
have been very helpful. Generous financial assistance from the Nordic Institute
of Theoretical Physics (Nordita) to MA is gratefully acknowledged.
\end{acknowledgements}


\begin{table*}[t]
\caption{Atomic data and derived abundances for Fe
lines using 3D and 2D convection simulations
\label{t:lines}
}
\begin{tabular}{lccccccccc} 
 \hline 
Species      &   Wavelength  &  log\,$gf$ &  $\chi_{\rm l}$ & 
$W_\lambda$ & log\,$\epsilon_{\rm Fe}$   & log\,$\epsilon_{\rm Fe}$ 
& log\,$\epsilon_{\rm Fe}$  & log\,$\epsilon_{\rm Fe}$  
& log\,$\epsilon_{\rm Fe}$$^{\rm a}$  \\
& [nm] & & [eV] & [pm] & ($200^2{\rm x}82$) &  ($100^2{\rm x}82$) & ($50^2{\rm x}82$) 
& ($50^2{\rm x}63$) & ($100{\rm x}82$) \\
\hline 
Fe\,{\sc i}  &       438.92451  &  -4.583  &   0.052  &    7.17  &    7.46  &    7.53  &    7.56  &   7.55 &   7.40  \\
             &       444.54717  &  -5.441  &   0.087  &    3.88  &    7.44  &    7.46  &    7.45  &   7.45 &   7.34  \\
             &       524.70503  &  -4.946  &   0.087  &    6.58  &    7.46  &    7.52  &    7.53  &   7.53 &   7.39  \\
             &       525.02090  &  -4.938  &   0.121  &    6.49  &    7.47  &    7.52  &    7.54  &   7.53 &   7.39  \\
             &       570.15444  &  -2.216  &   2.559  &    8.51  &    7.55  &    7.59  &    7.61  &   7.62 &   7.50  \\
             &       595.66943  &  -4.605  &   0.859  &    5.08  &    7.46  &    7.49  &    7.49  &   7.49 &   7.38  \\
             &       608.27104  &  -3.573  &   2.223  &    3.40  &    7.45  &    7.47  &    7.46  &   7.47 &   7.38  \\
             &       613.69946  &  -2.950  &   2.198  &    6.38  &    7.41  &    7.45  &    7.46  &   7.47 &   7.37  \\
             &       615.16182  &  -3.299  &   2.176  &    4.82  &    7.41  &    7.44  &    7.43  &   7.45 &   7.36  \\
             &       617.33354  &  -2.880  &   2.223  &    6.74  &    7.44  &    7.48  &    7.49  &   7.50 &   7.40  \\
             &       620.03130  &  -2.437  &   2.608  &    7.56  &    7.55  &    7.60  &    7.61  &   7.62 &   7.51  \\
             &       621.92808  &  -2.433  &   2.198  &    9.15  &    7.49  &    7.53  &    7.54  &   7.55 &   7.42  \\
             &       626.51338  &  -2.550  &   2.176  &    8.68  &    7.47  &    7.52  &    7.53  &   7.54 &   7.42  \\
             &       628.06182  &  -4.387  &   0.859  &    6.24  &    7.46  &    7.50  &    7.51  &   7.51 &   7.39  \\
             &       629.77930  &  -2.740  &   2.223  &    7.53  &    7.46  &    7.50  &    7.52  &   7.53 &   7.42  \\
             &       632.26855  &  -2.426  &   2.588  &    7.92  &    7.58  &    7.63  &    7.64  &   7.65 &   7.54  \\
             &       648.18701  &  -2.984  &   2.279  &    6.42  &    7.49  &    7.53  &    7.54  &   7.55 &   7.45  \\
             &       649.89390  &  -4.699  &   0.958  &    4.43  &    7.46  &    7.49  &    7.48  &   7.49 &   7.38  \\
             &       657.42285  &  -5.004  &   0.990  &    2.65  &    7.42  &    7.43  &    7.41  &   7.42 &   7.31  \\
             &       659.38706  &  -2.422  &   2.433  &    8.64  &    7.54  &    7.58  &    7.59  &   7.60 &   7.48  \\
             &       660.91104  &  -2.692  &   2.559  &    6.55  &    7.49  &    7.53  &    7.54  &   7.55 &   7.45  \\
             &       662.50220  &  -5.336  &   1.011  &    1.36  &    7.39  &    7.40  &    7.38  &   7.38 &   7.27  \\
             &       675.01523  &  -2.621  &   2.424  &    7.58  &    7.50  &    7.54  &    7.56  &   7.57 &   7.46  \\
             &       694.52051  &  -2.482  &   2.424  &    8.38  &    7.50  &    7.55  &    7.56  &   7.57 &   7.45  \\
             &       697.88516  &  -2.500  &   2.484  &    8.01  &    7.50  &    7.55  &    7.56  &   7.57 &   7.45  \\
             &       772.32080  &  -3.617  &   2.279  &    3.85  &    7.52  &    7.54  &    7.53  &   7.54 &   7.46  \\
             &                  &          &          &&$7.48\pm0.05$&$7.51\pm0.05$&$7.52\pm0.06$&$7.53\pm0.06$&$7.41\pm0.06$ \\
Fe\,{\sc ii} &       457.63334  &  -2.940  &   2.844  &    6.80  &    7.47  &    7.51  &    7.55  &   7.56 &   7.53   \\
             &       462.05129  &  -3.210  &   2.828  &    5.40  &    7.39  &    7.42  &    7.46  &   7.47 &   7.44  \\
             &       465.69762  &  -3.590  &   2.891  &    3.80  &    7.45  &    7.47  &    7.49  &   7.50 &   7.48  \\
             &       523.46243  &  -2.230  &   3.221  &    8.92  &    7.50  &    7.53  &    7.55  &   7.56 &   7.53  \\
             &       526.48042  &  -3.250  &   3.230  &    4.74  &    7.61  &    7.63  &    7.66  &   7.67 &   7.66  \\
             &       541.40717  &  -3.500  &   3.221  &    2.76  &    7.38  &    7.39  &    7.40  &   7.41 &   7.41  \\
             &       552.51168  &  -3.950  &   3.267  &    1.27  &    7.40  &    7.41  &    7.41  &   7.41 &   7.43  \\
             &       562.74892  &  -4.100  &   3.387  &    0.86  &    7.46  &    7.46  &    7.47  &   7.47 &   7.50  \\
             &       643.26757  &  -3.500  &   2.891  &    4.34  &    7.40  &    7.42  &    7.44  &   7.45 &   7.45  \\
             &       651.60767  &  -3.380  &   2.891  &    5.75  &    7.58  &    7.61  &    7.63  &   7.64 &   7.64  \\
             &       722.23923  &  -3.360  &   3.889  &    2.00  &    7.59  &    7.60  &    7.60  &   7.61 &   7.64  \\
             &       722.44790  &  -3.280  &   3.889  &    2.07  &    7.53  &    7.54  &    7.55  &   7.55 &   7.58  \\
             &       744.93305  &  -3.090  &   3.889  &    1.95  &    7.30  &    7.30  &    7.31  &   7.31 &   7.32  \\
             &       751.58309  &  -3.440  &   3.903  &    1.49  &    7.50  &    7.50  &    7.51  &   7.51 &   7.55  \\
             &       771.17205  &  -2.470  &   3.903  &    5.06  &    7.42  &    7.44  &    7.46  &   7.47 &   7.48  \\
             &                  &          &          &&$7.47\pm0.09$&$7.48\pm0.09$&$7.50\pm0.10$&$7.51\pm0.10$&$7.51\pm0.10$ \\
\hline  \\
\end{tabular}


\end{table*}

\end{document}